\documentclass[a4paper,11pt]{article}
\usepackage[DIV12]{typearea}
\usepackage{longtable}
\usepackage{booktabs,colortbl}
\usepackage{multirow}
\usepackage{amsmath}
\usepackage{amssymb}
\usepackage{bbm}
\usepackage[utf8]{inputenc}
\usepackage{pdflscape}
\usepackage{xspace}
\usepackage[caption=false]{subfig}
\usepackage{nicefrac}
\usepackage{graphicx}
\usepackage{cite}  
\usepackage[small]{caption2}
\usepackage{scalerel} %to scale subscripts in equations
\usepackage{mathtools}

\usepackage{adjustbox}%\adjincludegraphics
\usepackage{xfrac} % \sfrac for fractions that resemble x/y

\numberwithin{equation}{section}

\newcommand{\ttitle}{The three-dimensional $\mathcal{N} = 2$ superfishnet theory}

\newcommand{\SU}[1]{\ensuremath{\mathrm{SU}(#1)}}

\newcommand{\U}[1]{\ensuremath{\mathrm{U}(#1)}}

\newcommand{\e}{\mathrm{e}}
\newcommand{\I}{\mathrm{i}}

\newcommand{\x}{\ensuremath{\times}}

\newcommand{\dd}{\mathrm{d}}

\newcommand{\sumint}[1]{\mathrlap{\displaystyle\int_{#1}}\mathrlap{\textstyle\sum}}

\usepackage{xcolor}

\advance \headheight by 3.0truept       % for 12pt mandatory...
\usepackage{hyperref}
\hypersetup{
    pdftitle = {\ttitle},
    pdfauthor = {Kade}
}

\begin{document}

\begin{titlepage}

%\vspace*{-3.0cm}
\begin{flushright}
\normalsize{HU-EP-24/30-RTG}\\
\end{flushright}

\vspace*{1.0cm}

\begin{center}
{\Large\textbf{\boldmath \ttitle}\unboldmath}

\vspace{1cm}

\textbf{Moritz Kade}
\\[4mm]
\begin{small}
\texttt{\href{mailto:mkade@physik.hu-berlin.de}{mkade}@physik.hu-berlin.de}
\end{small}
\\[8mm]
\textit{\small Institut f\"{u}r Mathematik und Institut f\"{u}r Physik,\\
Humboldt-Universit\"{a}t zu Berlin,\\
Zum Großen Windkanal 2, 12489 Berlin, Germany}
\end{center}

\vspace{1cm}

\vspace*{1.0cm}

\begin{abstract}
We propose a double-scaling limit of $\beta$-deformed ABJM theory in three-dimensional $\mathcal{N} = 2$ superspace, and a non-local deformation thereof.
Due to the regular appearance of the theory's Feynman supergraphs, we refer to this superconformal and integrable theory as the superfishnet theory.
We use techniques inspired by the integrability of bi-scalar fishnet theory and adapted to superspace to calculate the zero-mode-fixed thermodynamic free energy, the corresponding critical coupling, and the exact all-loop scaling dimensions of various operators.
Furthermore, we confirm the results of the supersymmetric dynamical fishnet theory by applying our methods to four-dimensional $\mathcal{N} = 1$ superspace.
\end{abstract}

\end{titlepage}

\newpage

%%%%%%%%%%%%%%%%%%%%%%%%%%%%%%%%%%%%%%%%%%%%%%%%%%%%%%%%%%%%%%%%%%%%%%%%%%%%%%%%%%%%%%%%%%%%%%%%%%%%%%%%%%%%%%%%%%%%%%%%%%%%%%%%%%%%%%%%%
\section{Introduction}
\label{sec:Introduction}

In three dimensions, superconformal field theories (SCFT) have been extensively explored, in large part owing to the development of the ABJM theory, an $\mathcal{N} = 6$ supersymmetric Chern-Simons theory, which describes the low-energy effective action of $\mathrm{N}$ coincident M2-branes probing a \(\mathbb{C}^4/\mathbb{Z}_k\) singularity \cite{Aharony:2008ug,Benna:2008zy}. 
Its large-$\mathrm{N}$ limit is dual to string theory on $\mathrm{AdS}_4 \times \mathbb{CP}^3$ via the AdS/CFT correspondence. 
Furthermore, its integrability \cite{Beisert:2010jr,Klose:2010ki} allows us to find anomalous dimensions in the planar limit by mapping the problem to the spectral problem of an integrable spin chain \cite{Minahan:2008hf,Gaiotto:2008cg,Gromov:2008qe,Zwiebel:2009vb,Minahan:2009te,Minahan:2009wg,Cavaglia:2014exa}.

In the study of $\mathcal{N}=4$ super Yang-Mills theory (SYM), it is helpful to consider deformations in the double-scaling limit \cite{Gurdogan:2015csr}.
On the one hand, these deformations appear integrable on a level of the Feynman graph due to their simple, regular patterns.
Among others \cite{Kazakov:2022dbd,Pittelli:2019ceq,Mamroud:2017uyz,Kade:2023xet}, a prominent example is the bi-scalar fishnet theory \cite{Gurdogan:2015csr,Kazakov:2018qbr}, which admits fishnet-like Feynman graphs in the perturbative expansion.
Similar deformations of ABJM were considered in \cite{Caetano:2016ydc}, where the Bethe Ansatz solved the spectral problem.
On the other hand, these theories have fewer degrees of freedom and are non-unitary.
Lifting some of the deformations reinstalls some degrees of freedom for the price of seemingly less regular Feynman graphs; this is the case for the dynamical fishnet theory \cite{Kazakov:2018gcy}.
Recently, for the supersymmetric double-scaled deformations, the superspace formalism, where one supergraph collectively captures many ordinary Feynman graphs, has shown that the supergraphs of double-scaled $\beta$-deformations of $\mathcal{N} =4$ have still a regular brick wall pattern \cite{Kade:2024ucz}.
Working in a supersymmetric covariant way allows techniques developed for the bi-scalar fishnet theory to be transferred to the super brick wall theory with more degrees of freedom, three scalars and three fermions.

In this paper, we adopt the findings of \cite{Kade:2024ucz} to ABJM and present a new SCFT in three dimensions characterized by a fishnet pattern in its Feynman supergraphs.
Hence, we refer to it as the superfishnet theory.
It emerges from a particular double-scaling limit of the ABJM theory, which involves both the 't Hooft coupling $\lambda$ and the deformation parameter of the $\beta$-deformation of the theory. 
The $\beta$-deformation of ABJM \cite{Imeroni:2008cr,He:2013hxd,Caetano:2016ydc} breaks some of the supersymmetry and modifies the structure of the theory's operator spectrum. 
Our interest lies in a regime where the deformation parameter and 't Hooft coupling are both taken in a correlated manner to infinity and zero, respectively, while keeping a particular combination of the parameters finite.

This double-scaling limit decouples the gauge degrees of freedom and drastically simplifies the structure of perturbative calculations, since only two of the original interaction vertices of the $\beta$-deformed ABJM survive the limit.
The obtained theory is non-unitary, since the double-scaling limit removes the hermitian conjugated counterparts from the theory.
We expect the theory to be integrable, but we lack a superconformal graph-builder that satisfies a Yang-Baxter equation, as in the bosonic fishnet theory \cite{Gromov:2017cja}.
Nevertheless, we obtain the theory from ABJM theory, which is assumed to be integrable in the planar limit.
However, similar to ordinary, bosonic Feynman graphs, the supergraphs' regular structure makes it possible to determine the critical coupling and anomalous dimensions of some operators exactly. 
For the former, we employ the method of inversion relations \cite{Zamolodchikov:1980mb,Kade:2023xet}, initially developed in statistical physics \cite{StroganovInvRelLatticeModels,Baxter:1982xp,Baxter:1989ct,Baxter:2002qg,Pokrovsky_1982,bousquetmelou1999inversion}, to the three-dimensional vacuum supergraphs.
For the latter, we generalize the method of identifying repeating graph-builders in the weak-coupling expansion of specific correlation functions \cite{Gromov:2018hut} and then solve an eigenvalue problem for them.
This gives us access to the anomalous dimension of the operator corresponding to the eigenfunction.

The structure of this paper is as follows. Section 2 provides a detailed review of the ABJM theory and its $\beta$-deformation in a supersymmetry covariant manner.
We perform the double-scaling limit and obtain the action of the superfishnet theory as a three-dimensional $\mathcal{N} = 2$ superspace integral.
Section 3 also introduces a non-local deformation of the superfishnet theory, the generalized superfishnet model.
Furthermore, the super Feynman rules are derived, and we establish auxiliary integral relations on superspace, notably the super chain rules and the super x-unity relation.  
In Section 4, we use the latter to calculate the critical coupling of the generalized superfishnet theory using the method of inversion relations.
Section 5 explores the weak-coupling expansion of some correlation functions and uses the regular nature of the supergraphs to calculate the exact anomalous dimension of exchanged operators by diagonalizing of supergraph-builders.
We conclude in section 6 with a summary of our findings and an outlook on future research directions.
Two appendices provide details on the notations used throughout the paper, as well as details on the calculations of the superspace integral relations, the diagonalization of the graph-builders, and an application to the super brick wall theory.

%In summary, the new SCFT presented here offers a fascinating extension of three-dimensional superconformal theories, combining the rich structure of ABJM theory with novel computational tools afforded by the double-scaling limit. The exact results derived from its regular Feynman supergraphs open the door to further studies in both the non-perturbative dynamics of SCFTs and potential applications to holography and integrability.

%%%%%%%%%%%%%%%%%%%%%%%%%%%%%%%%%%%%%%%%%%%%%%%%%%%%%%%%%%%%%%%%%%%%%%%%%%%%%%%%%%%%%%%%%%%%%%%%%%%%%%%%%%%%%%%%%%%%%%%%%%%%%%%%%%%%%%%%%
\section{The double-scaled $\boldsymbol{\beta}$-deformation of ABJM}
\label{sec:DoubleScaledBetaDeformationOfABJM}
We review the $\mathcal{N}=2$ superspace formulation of three-dimensional, $\mathcal{N}=6$ ABJM theory \cite{Aharony:2008ug,Benna:2008zy,Leoni:2010tb,Bianchi:2012cq}, following closely \cite{Benna:2008zy}, and show the reduction to the superfishnet theory by combining R-symmetry deformations with double-scaling limits.
Double-scaled deformations of the R-symmetry of the ABJM theory, supersymmetric and non-supersymmetric, were studied in \cite{Caetano:2016ydc}.
However, this section focuses on the $\beta$-deformation, which preserves $\mathcal{N}=2$ supersymmetry, which allows us to perform the deformation at the level of the superspace action \cite{Imeroni:2008cr}.
In a second step, we take a double-scaling limit of the 't Hooft coupling and the deformation parameter, which renders the theory non-unitary, but leaves us with a set of very restricted, quartic interaction terms.
We present the super-Feynman rules for the resulting $\mathcal{N}=2$ superfishnet theory.

\subsection{\boldmath ABJM $\U{\mathrm{N}} \otimes \U{\mathrm{N}}$ gauge theory}
The ABJM theory is an $\mathcal{N}=6$ superconformal Chern-Simons theory of the gauge group $\U{\mathrm{N}} \otimes \U{\mathrm{N}}$ with levels $k$ and $-k$ and bi-fundamental matter, which couples to the gauge fields, and a superpotential.
Its global superconformal algebra is $\mathfrak{osp}(6 \vert 4)$.
Schematically, we write the action as \cite{Benna:2008zy}
\begin{equation}
S_\mathrm{ABJM} = -\I\, \tfrac{k}{\lambda} \cdot S_\mathrm{CS} + S_\mathrm{mat} + \tfrac{\lambda}{k} \cdot S_\mathrm{pot} ~.
\label{eq:Action_ABJM}
\end{equation}
It depends on the Chern-Simons level $k$ and the 't Hooft coupling $\lambda = \frac{\mathrm{N}}{k}$.
The action can be conveniently formulated in three-dimensional $\mathcal{N}=2$ superspace.
We present its different terms along with the corresponding field content, following the notations of \cite{Benna:2008zy}.
\begin{itemize}
\item
The Chern-Simons part $S_\mathrm{CS}$ is given by 
\begin{equation}
S_\mathrm{CS} \left[ \mathcal{V} , \hat{\mathcal{V}} \right]
=
\int \dd^3 x \; \dd^2 \theta \, \dd^2 \bar{\theta}
\int_0^1 \dd t ~
\mathrm{tr}
\left[
\mathcal{V} \bar{D}^\alpha 
\left(
\e^{t\mathcal{V}} D_\alpha \e^{-t\mathcal{V}}
\right)
-
\hat{\mathcal{V}} \bar{D}^\alpha 
\left(
\e^{t\hat{\mathcal{V}}} D_\alpha \e^{-t\hat{\mathcal{V}}}
\right)
\right] ~.
\end{equation}
Its degrees of freedom are two gauge vector superfields $\mathcal{V}$ and $\hat{\mathcal{V}}$, they correspond to the representations $\left( \mathbf{N}\otimes \bar{\mathbf{N}} , \mathbf{1} \right)$ and $\left( \mathbf{1} , \mathbf{N}\otimes \bar{\mathbf{N}} \right)$ of the gauge group $\U{\mathrm{N}} \otimes \U{\mathrm{N}}$, respectively.
The supersymmetric covariant derivatives $D_\alpha$ and $\bar{D}_\alpha$ are defined in \eqref{eq:SuperCovariantDerivatives}.
The vector superfield contains the gauge field $A_\mu$ and admits an expansion
\begin{equation}
\mathcal{V}
=
2\I\, \theta \bar{\theta}\, \sigma (x) 
+ 2\, \theta \gamma^\mu \bar{\theta} \, A_\mu (x) 
+ \sqrt{2} \I\, \theta^2\, \bar{\theta}\bar{\chi} (x)
- \sqrt{2} \I\, \bar{\theta}^2\, \theta \chi (x)
+ \theta^2\, \bar{\theta}^2\, \mathrm{D} (x)
\end{equation}
in Gra\ss mann components.
Analogously, $\hat{\mathcal{V}}$ contains $\hat{A}_\mu$.
The remaining component-fields are auxiliary scalars ($\sigma$, $\mathrm{D}$ and their hatted versions) and auxiliary fermions ($\chi$, $\bar{\chi}$ and their hatted versions).

\item
The matter part $S_\mathrm{mat}$ of the ABJM theory is 
\begin{equation}
S_\mathrm{mat} \left[ \mathcal{Z}, \mathcal{W}, \bar{\mathcal{Z}}, \bar{\mathcal{W}} , \mathcal{V}, \hat{\mathcal{V}} \right]
=
\int \dd^3 x \; \dd^2 \theta \, \dd^2 \bar{\theta} ~
\mathrm{tr}
\left[
- \bar{\mathcal{Z}}_A \e^{-\mathcal{V}} \mathcal{Z}^A \e^{\hat{\mathcal{V}}}
- \bar{\mathcal{W}}^A \e^{-\hat{\mathcal{V}}} \mathcal{W}_A \e^{\mathcal{V}}
\right] ~.
\end{equation}
It contains the chiral superfields $\mathcal{Z}$ and $\mathcal{W}$, as well as the anti-chiral superfields $\bar{\mathcal{Z}}$ and $\bar{\mathcal{W}}$ as matter degrees of freedom.
Further, it includes the coupling of the matter fields to the gauge fields $\mathcal{V}$ and $\hat{\mathcal{V}}$.
The chiral superfields $\mathcal{Z}$ and $\mathcal{W}$ transform in the representations $(\mathbf{2},\mathbf{1})$ and $(\mathbf{1},\bar{\mathbf{2}})$ of the global symmetry group $\SU{2}\otimes \SU{2}$, respectively.
We denote the corresponding indices by capital Latin letters.  
Furthermore, they transform in the bi-fundamental representations $(\mathbf{N}, \bar{\mathbf{N}})$ and $(\bar{\mathbf{N}}, \mathbf{N})$ of the gauge group $\U{\mathrm{N}} \otimes \U{\mathrm{N}}$, respectively.
The anti-chiral superfields $\bar{\mathcal{Z}}$ and $\bar{\mathcal{W}}$ transform accordingly in the representations $(\bar{\mathbf{2}},\mathbf{1})$ and $(\mathbf{1}, \mathbf{2})$ of the global symmetry group and in the representations $(\bar{\mathbf{N}}, \mathbf{N})$ and $(\mathbf{N}, \bar{\mathbf{N}})$ of the gauge group, respectively.
The expansions of the chiral superfields in Gra\ss mann components are
\begin{subequations}
\begin{align}
\mathcal{Z}^A 
&= Z^A (x_+) + \sqrt{2}\, \theta \zeta^A (x_+) + \theta^2\, F^A (x_+) ~,\\
\mathcal{W}_A
&= W_A (x_+) + \sqrt{2}\, \theta \omega_A (x_+) + \theta^2\, G_A (x_+) ~,
\end{align}
\label{eq:ComponentExpansion_ChiralSuperfields}%
\end{subequations}
and the ones of the anti-chiral superfields are 
\begin{subequations}
\begin{align}
\bar{\mathcal{Z}}_A 
&= Z^\dagger_A (x_-) - \sqrt{2}\, \bar{\theta} \zeta^\dagger_A (x_-) - \bar{\theta}^2\, F^\dagger_A (x_-) ~,\\
\bar{\mathcal{W}}^A
&= W^{\dagger A} (x_-) - \sqrt{2}\, \bar{\theta} \omega^{\dagger A} (x_-) - \bar{\theta}^2 G^{\dagger A} (x_-) ~.
\end{align}
\label{eq:ComponentExpansion_AntiChiralSuperfields}%
\end{subequations}
The bosons $F^A$, $G_A$, $F^\dagger_A$ and $G^{\dagger A}$ are auxiliary fields and we introduced the shorthand notation $x^\mu_\pm = x^\mu \pm \I \theta \gamma^\mu \bar{\theta}$.

\item
The matter superfields couple in a quartic interaction in the superpotential action
\begin{equation}
\begin{split}
S_\mathrm{pot} \left[ \mathcal{Z}, \mathcal{W}, \bar{\mathcal{Z}}, \bar{\mathcal{W}} \right]
=
& \int \dd^3 x \; \dd^2 \theta ~
\tfrac{1}{4}\, \varepsilon_{AC}\varepsilon^{BD}\;
\mathrm{tr}
\left[
\mathcal{Z}^A \mathcal{W}_B \mathcal{Z}^C \mathcal{W}_D
\right]\\
& +
\int \dd^3 x \; \dd^2 \bar{\theta} ~
\tfrac{1}{4}\, \varepsilon^{AC}\varepsilon_{BD}\;
\mathrm{tr}
\left[
\bar{\mathcal{Z}}_A \bar{\mathcal{W}}^B \bar{\mathcal{Z}}_C \bar{\mathcal{W}}^D
\right] ~.
\end{split}
\label{eq:Superpotential_ABJM}
\end{equation}
Here we denote the antisymmetric tensor of $\SU{2}$ by $\varepsilon$ with the conventions $\varepsilon^{12} = \varepsilon_{21} = +1$.
The superpotential action and the other parts of \eqref{eq:Action_ABJM} are symmetric under a global $\SU{2} \otimes \SU{2}$, as well as under a global $\U{1}_\mathrm{R}$ symmetry, which is the manifest part of the R-symmetry in a three-dimensional $\mathcal{N}=2$ superspace action.
The $\U{1}_\mathrm{R}$ acts on fermionic coordinates as $\theta \rightarrow \e^{\I \alpha}\theta$ and $\bar{\theta} \rightarrow \e^{-\I \alpha} \bar{\theta}$, and on the matter fields as
\begin{subequations}
\begin{align}
\mathcal{Z}^A &\rightarrow \e^{-\frac{\I}{2} \alpha}\, \mathcal{Z}^A (x,\e^{\I \alpha}\theta ,\e^{-\I \alpha} \bar{\theta}) ~,
&&\mathcal{W}_A \rightarrow \e^{-\frac{\I}{2} \alpha}\, \mathcal{W}_A (x,\e^{\I \alpha}\theta ,\e^{-\I \alpha} \bar{\theta}) ~,  \\
\bar{\mathcal{Z}}_A &\rightarrow \e^{\frac{\I}{2} \alpha}\, \bar{\mathcal{Z}}_A (x,\e^{\I \alpha}\theta ,\e^{-\I \alpha} \bar{\theta}) ~,
&&\bar{\mathcal{W}}^A \rightarrow \e^{\frac{\I}{2} \alpha}\, \bar{\mathcal{W}}^A (x,\e^{\I \alpha}\theta ,\e^{-\I \alpha} \bar{\theta}) ~.
\end{align}
\label{eq:U1_intakt_transformation}%
\end{subequations}
The global symmetry $\SU{2} \otimes \SU{2}$ enhances the manifest R-symmetry $\U{1}_\mathrm{R}$ to the R-symmetry $\SU{4}_\mathrm{R}$ of $\mathcal{N}=6$ ABJM theory \eqref{eq:Action_ABJM}.
To summarize, we display the charges of the degrees of freedom under the three Cartan-generators of $\mathfrak{su}(4)_\mathrm{R}$ in table \ref{tab:Rcharges}.
\end{itemize}

%\begin{table}[]
%\begin{tabular}{ll|llll|llll|ll|}
%\cline{3-12}
%                      &  & $\mathcal{Z}^1$ & $\mathcal{Z}^2$ & $\mathcal{W}_1$ & $\mathcal{W}_2$ & $\bar{\mathcal{Z}}_1$ & $\bar{\mathcal{Z}}_2$ & $\bar{\mathcal{W}}^1$ & $\bar{\mathcal{W}}^2$ & $\mathcal{V}$ & $\hat{\mathcal{V}}$ \\ \hline
%\multicolumn{1}{|l}{} $\mathfrak{su}(2)$ & $q^1$ & $\phantom{-}\frac{1}{2}$ & $-\frac{1}{2}$ & $\phantom{-}0$ & $\phantom{-}0$ & $-\frac{1}{2}$ & $\phantom{-}\frac{1}{2}$ & $\phantom{-}0$ & $\phantom{-}0$ & $0$ & $0$ \\
%\multicolumn{1}{|l}{} $\mathfrak{su}(2)$ & $q^2$ & $\phantom{-}0$ & $\phantom{-}0$ & $-\frac{1}{2}$ & $\phantom{-}\frac{1}{2}$ & $\phantom{-}0$ & $\phantom{-}0$ & $\phantom{-}\frac{1}{2}$ & $-\frac{1}{2}$ & $0$ & $0$ \\
%\multicolumn{1}{|l}{} $\mathfrak{u}(1)$ & $q^3$ & $\phantom{-}\frac{1}{2}$ & $\phantom{-}\frac{1}{2}$ & $-\frac{1}{2}$ & $-\frac{1}{2}$ & $-\frac{1}{2}$ & $-\frac{1}{2}$ & $\phantom{-}\frac{1}{2}$ & $\phantom{-}\frac{1}{2}$ & $0$ & $0$ \\ \hline
%\end{tabular}
%\centering
%\end{table}
\begin{table}[]
\centering
\begin{tabular}{lc|llll|llll|ll}
 &  & $\mathcal{Z}^1$ & $\mathcal{Z}^2$ & $\mathcal{W}_1$ & $\mathcal{W}_2$ & $\bar{\mathcal{Z}}_1$ & $\bar{\mathcal{Z}}_2$ & $\bar{\mathcal{W}}^1$ & $\bar{\mathcal{W}}^2$ & $\mathcal{V}$ & $\hat{\mathcal{V}}$ \\ \hline
$\mathfrak{su}(2)$ & $q^1$ & $\phantom{-}\frac{1}{2}$ & $-\frac{1}{2}$ & $\phantom{-}0$ & $\phantom{-}0$ & $-\frac{1}{2}$ & $\phantom{-}\frac{1}{2}$ & $\phantom{-}0$ & $\phantom{-}0$ & $0$ & $0$ \\
$\mathfrak{su}(2)$ & $q^2$ & $\phantom{-}0$ & $\phantom{-}0$ & $-\frac{1}{2}$ & $\phantom{-}\frac{1}{2}$ & $\phantom{-}0$ & $\phantom{-}0$ & $\phantom{-}\frac{1}{2}$ & $-\frac{1}{2}$ & $0$ & $0$ \\
$\mathfrak{u}(1)$ & $R = q^3$ & $\phantom{-}\frac{1}{2}$ & $\phantom{-}\frac{1}{2}$ & $\phantom{-}\frac{1}{2}$ & $\phantom{-}\frac{1}{2}$ & $-\frac{1}{2}$ & $-\frac{1}{2}$ & $-\frac{1}{2}$ & $-\frac{1}{2}$ & $0$ & $0$
\end{tabular}
\caption{The charges of the superfields in  the ABJM theory under the Cartan-generators of the subalgebras of the R-symmetry $\mathfrak{su}(4)_\mathrm{R}$. The $\beta$-deformation breaks the R-symmetry, only keeping the $\mathfrak{u}(1)$ component intact.}
\label{tab:Rcharges}
\end{table}

In order to establish the genus expansion in the large-$\mathrm{N}$ limit, we find it convenient to rescale the (anti-)chiral superfields as $(\mathcal{Z}, \mathcal{W}, \bar{\mathcal{Z}}, \bar{\mathcal{W}}) \rightarrow \sqrt{\mathrm{N}} \cdot (\mathcal{Z}, \mathcal{W}, \bar{\mathcal{Z}}, \bar{\mathcal{W}})$ and the gauge vector superfields as $(\mathcal{V}, \hat{\mathcal{V}}) \rightarrow \lambda \cdot (\mathcal{V}, \hat{\mathcal{V}})$.
This way, the ABJM action \eqref{eq:Action_ABJM} turns into 
%$S_\mathrm{ABJM} \rightarrow S'_\mathrm{ABJM}$ with
\begin{equation}
S'_\mathrm{ABJM}
=
- \I\, \tfrac{k}{\lambda} \cdot S_\mathrm{CS} \left[ \lambda \mathcal{V} ,\lambda \hat{\mathcal{V}} \right]
+ \mathrm{N} \cdot S_\mathrm{mat}\left[ \mathcal{Z}, \mathcal{W}, \bar{\mathcal{Z}}, \bar{\mathcal{W}}, \lambda \mathcal{V} ,\lambda \hat{\mathcal{V}} \right]
+ \mathrm{N}\lambda^2 \cdot S_\mathrm{pot} \left[ \mathcal{Z}, \mathcal{W}, \bar{\mathcal{Z}}, \bar{\mathcal{W}} \right] ~.
\label{eq:Action_ABJM_prime}
\end{equation}
The action $S'_\mathrm{ABJM}$ will be subject to the $\beta$-deformation and a double-scaling limit, eventually yielding the superfishnet theory.

\subsection{\boldmath $\beta$-Deformation and double-scaling limit}
We obtain the superfishnet theory from the action \eqref{eq:Action_ABJM_prime} by a two-step procedure, consisting of a $\beta$-deformation of the $\SU{4}_\mathrm{R}$ R-symmetry and a consecutive double-scaling limit of the 't Hooft coupling $\lambda$ and the deformation parameter.
\begin{enumerate}
\item 
The $\beta$-deformation of the ABJM theory on the level of the superpotential was described in \cite{Imeroni:2008cr} and on the level of the component fields in \cite{Caetano:2016ydc,He:2013hxd,Chen:2016geo}.
It consists of twisting the two $\U{1}$ Cartan-subgroups in the $\SU{2} \otimes \SU{2}$ subgroup of the global R-symmetry group $\SU{4}_\mathrm{R}$.
Therefore, the $\U{1}_\mathrm{R}$-factor corresponding to the transformations \eqref{eq:U1_intakt_transformation} stays intact and will be the residual R-symmetry of the $\beta$-deformed theory, which hence is $\mathcal{N}=2$ supersymmetric.
This is important, since it allows us to keep our $\mathcal{N}=2$ superspace formalism untouched by the deformation.

In the superspace action $S'_\mathrm{ABJM}$ in \eqref{eq:Action_ABJM_prime}, the deformation is implemented by replacing ordinary products of superfields by 
\begin{equation}
\Phi_1 \cdots \Phi_p
~ \rightarrow ~
\e^{-\frac{\I}{2} \sum^p_{m > n} \beta \cdot \varepsilon_{ij}\, q^i_{\Phi_m} q^j_{\Phi_n}}\;
\Phi_1 \cdots \Phi_p ~.
\label{eq:BetaDeformation_StarProduct}
\end{equation}
Here, the antisymmetric $\varepsilon_{ij}$ has the component $\varepsilon_{12} = 1$ and the charges $q^1_\Phi$ and $q^2_\Phi$ correspond to the charges of the field $\Phi$ under the Cartan-generators of the $\SU{2} \otimes \SU{2}$ subgroup.
They are listed in the first two rows of table \ref{tab:Rcharges}.
We denote the deformation parameter as $\beta$.

Since the gauge vector superfields do not transform under the R-symmetry, see table \ref{tab:Rcharges}, we find the Chern-Simons part of the action \eqref{eq:Action_ABJM_prime} unaltered by the $\beta$-deformation $S_\mathrm{CS} \rightarrow S_\mathrm{CS}$.
The matter superfields $\mathcal{Z}$ and $\bar{\mathcal{Z}}$, as well as $\mathcal{W}$ and $\bar{\mathcal{W}}$ have opposite charges under the R-symmetry, respectively, resulting in an unchanged matter term of the action \eqref{eq:Action_ABJM_prime}, $S_\mathrm{mat} \rightarrow S_\mathrm{mat}$.

The only term affected by the deformation in the action \eqref{eq:Action_ABJM_prime} is the superpotential part $S_\mathrm{pot}$.
The holomorphic and anti-holomorphic superpotentials \eqref{eq:Superpotential_ABJM} get deformed as
\begin{subequations}
\begin{align}
\mathrm{tr} 
\left[
\mathcal{Z}^1 \mathcal{W}_2 \mathcal{Z}^2 \mathcal{W}_1
-
\mathcal{Z}^1 \mathcal{W}_1 \mathcal{Z}^2 \mathcal{W}_2
\right]
\rightarrow
\mathrm{tr} 
\left[
q \cdot \mathcal{Z}^1 \mathcal{W}_2 \mathcal{Z}^2 \mathcal{W}_1
-
q^{-1} \cdot \mathcal{Z}^1 \mathcal{W}_1 \mathcal{Z}^2 \mathcal{W}_2
\right] ~,\\
\mathrm{tr} 
\left[
\bar{\mathcal{Z}}_1 \bar{\mathcal{W}}^2 \bar{\mathcal{Z}}_2 \bar{\mathcal{W}}^1
-
\bar{\mathcal{Z}}_1 \bar{\mathcal{W}}^1 \bar{\mathcal{Z}}_2 \bar{\mathcal{W}}^2
\right]
\rightarrow
\mathrm{tr} 
\left[
q \cdot \bar{\mathcal{Z}}_1 \bar{\mathcal{W}}^2 \bar{\mathcal{Z}}_2 \bar{\mathcal{W}}^1
-
q^{-1} \cdot \bar{\mathcal{Z}}_1 \bar{\mathcal{W}}^1 \bar{\mathcal{Z}}_2 \bar{\mathcal{W}}^2
\right] ~,
\end{align}
\end{subequations}
respectively \cite{Imeroni:2008cr}.
We use the abbreviation $q = \e^{\frac{\I}{4} \beta}$, not to be confused with the charges.

To summarize, we find the action of the $\beta$-deformation of $S'_\mathrm{ABJM}$ to be
\begin{equation}
\begin{split}
S'_{\beta ,\, \mathrm{ABJM}}
=
& - \I\, \tfrac{k}{\lambda} \cdot S_\mathrm{CS} \left[ \lambda \mathcal{V} ,\lambda \hat{\mathcal{V}} \right]
+ \mathrm{N} \cdot S_\mathrm{mat}\left[ \mathcal{Z}, \mathcal{W}, \bar{\mathcal{Z}}, \bar{\mathcal{W}}, \lambda \mathcal{V} ,\lambda \hat{\mathcal{V}} \right]\\
& + \mathrm{N}\lambda^2 
\int \dd^3 x \; \dd^2 \theta ~
\tfrac{1}{2}\,
\mathrm{tr} 
\left[
q \cdot \mathcal{Z}^1 \mathcal{W}_2 \mathcal{Z}^2 \mathcal{W}_1
-
q^{-1} \cdot \mathcal{Z}^1 \mathcal{W}_1 \mathcal{Z}^2 \mathcal{W}_2
\right]\\
& + \mathrm{N}\lambda^2 
\int \dd^3 x \; \dd^2 \bar{\theta} ~
\tfrac{1}{2}\,
\mathrm{tr} 
\left[
q \cdot \bar{\mathcal{Z}}_1 \bar{\mathcal{W}}^2 \bar{\mathcal{Z}}_2 \bar{\mathcal{W}}^1
-
q^{-1} \cdot \bar{\mathcal{Z}}_1 \bar{\mathcal{W}}^1 \bar{\mathcal{Z}}_2 \bar{\mathcal{W}}^2
\right] ~.
\end{split}
\label{eq:Action_ABJM_prime_betaDef}
\end{equation}

\item
The second step consists of taking the so-called double-scaling limit of the 't Hooft coupling $\lambda$ and the deformation parameter $q$ of the $\beta$-deformed ABJM theory \eqref{eq:Action_ABJM_prime_betaDef} \cite{Caetano:2016ydc}.
We consider the limit $\lambda \rightarrow 0$ and $q \rightarrow \infty$ (or equivalently $\beta \rightarrow - \I \infty$), while requiring the product $2\xi := q \cdot \lambda^2$ to stay finite.
The double-scaling procedure has two consequences.

Firstly, due to the limit $\lambda \rightarrow 0$, the matter part of the action $S'_{\beta ,\, \mathrm{ABJM}}$ in \eqref{eq:Action_ABJM_prime_betaDef} loses its dependency on the gauge fields $\mathcal{V}$ and $\hat{\mathcal{V}}$.
We find $S_\mathrm{mat} \rightarrow S_\mathrm{mat}\left[ \mathcal{Z}, \mathcal{W}, \bar{\mathcal{Z}}, \bar{\mathcal{W}}, 0 ,0 \right]$.
Hence, the gauge vector superfields decouple.
We focus on the remainder of the action, which contains solely the matter superfields, discarding the Chern-Simons action $S_\mathrm{CS}$ in the following.

Secondly, the terms proportional to $q^{-1}$ will vanish in the double-scaling limit because $q^{-1} \rightarrow 0$, and there is no factor to compensate.
However, the terms proportional to $q$ combine with the prefactor $\lambda^2$ to the new coupling $\xi$, which we require to be finite.

We could also have conversely considered the limit $q \rightarrow 0$, while keeping the product $q^{-1}\lambda^2$ fixed.
In this way, we would keep the other two terms in the superpotential part.
However, the obtained theory would be the hermitian conjugate of the double-scaled theory with $q \rightarrow \infty$ and $\xi$ fixed.
\end{enumerate}
Finally, we find the action
\begin{equation}
\begin{split}
S_{\mathrm{SFN}}
~=~
&\mathrm{N}
\int \dd^3 x \; \dd^2 \theta \, \dd^2 \bar{\theta} ~
\mathrm{tr}
\left[
- \bar{\mathcal{Z}}_A \mathcal{Z}^A
- \bar{\mathcal{W}}^A \mathcal{W}_A
\right]\\
& + \mathrm{N}\xi 
\int \dd^3 x \; \dd^2 \theta ~
\mathrm{tr} 
\left[
´\mathcal{Z}^1 \mathcal{W}_2 \mathcal{Z}^2 \mathcal{W}_1
\right]
+ \mathrm{N}\xi 
\int \dd^3 x \; \dd^2 \bar{\theta} ~
\mathrm{tr} 
\left[
\bar{\mathcal{Z}}_1 \bar{\mathcal{W}}^2 \bar{\mathcal{Z}}_2 \bar{\mathcal{W}}^1
\right]
\end{split}
\label{eq:Action_superfishnet_Z_W}
\end{equation}
for the double-scaled $\beta$-deformation of ABJM theory, defining what we call superfishnet theory.
It is a three-dimensional $\mathcal{N}=2$ supersymmetric field theory of four chiral and four anti-chiral superfields.
The double-scaling limit broke unitarity; however, similarly to the double-scaling limits of $\mathcal{N}=4$ SYM, this sacrifice will lead to very regular Feynman supergraphs, as shown below.
Since the $\beta$-deformation breaks the global $\SU{2} \otimes \SU{2}$ group, the symmetry algebra is reduced to the superconformal $\mathfrak{osp}(2 \vert 4)$ algebra.

\subsection{Component action}
Let us derive the superfishnet action \eqref{eq:Action_superfishnet_Z_W} in bosonic spacetime.
The strategy is to replace the superfields in \eqref{eq:Action_superfishnet_Z_W} by their expansions in Gra\ss mann components \eqref{eq:ComponentExpansion_ChiralSuperfields} and \eqref{eq:ComponentExpansion_AntiChiralSuperfields}.
Then we can eliminate the remaining bosonic auxiliary fields $F^A$, $G_A$, $F_A^\dagger$ and $G^{\dagger A}$ by using their equations of motion\footnote{We have used identities like $\theta \varphi \cdot \theta \psi = - \frac{1}{2} \theta^2 \varphi \psi$ and $\bar{\theta} \bar{\varphi} \cdot \bar{\theta} \bar{\psi} = - \frac{1}{2} \bar{\theta}^2 \bar{\varphi} \bar{\psi}$, see appendix \ref{sec:Notations} for our conventions. We also do not distinguish the notations $\zeta_A^{\dagger \alpha}$ and $\bar{\zeta}_A^{\alpha}$ for fermions.}
\begin{equation}
\begin{aligned}[c]
F^1 &= \xi \, W^{\dagger 2} Z_2^\dagger W^{\dagger 1} ~,\\
F^2 &= \xi \, W^{\dagger 1} Z_1^\dagger W^{\dagger 2} ~,\\
\end{aligned}
\quad
\begin{aligned}[c]
G_1 &= \xi \, Z_1^{\dagger} W^{\dagger 2} Z_2^{\dagger} ~,\\
G_2 &= \xi \, Z_2^{\dagger} W^{\dagger 1} Z_1^{\dagger} ~,\\
\end{aligned}
\quad
\begin{aligned}[c]
F_1^\dagger &= - \xi \, W_2 Z^2 W_1 ~,\\
F_2^\dagger &= - \xi \, W_1 Z^1 W_2 ~,\\
\end{aligned}
\quad
\begin{aligned}[c]
G^{\dagger 1} &= - \xi \, Z^1 W_2 Z^2 ~,\\
G^{\dagger 2} &= - \xi \, Z^2 W_1 Z^1 ~.\\
\end{aligned}
\end{equation}
Eventually, we find the superfishnet action formulated on bosonic spacetime to be
\begin{equation}
\begin{split}
& S_\mathrm{SFN}
=
\mathrm{N}
\int \dd^3 x ~
\mathrm{tr}
\left\lbrace
Z_A^\dagger \square Z^A
+ W^{\dagger A} \square W_A
+ \I\, \zeta_A^{\dagger \alpha} \gamma^\mu_{\alpha \beta} \partial_\mu \zeta^{A \beta}
+ \I\, \omega^{\dagger A \alpha} \gamma^\mu_{\alpha \beta} \partial_\mu \omega_A^{\beta}
\right. \\
&+ \xi^2
\left[
Z^1 W_2 Z^2 Z_1^\dagger W^{\dagger 2} Z_2^\dagger +
W_1 Z^1 W_2 W^{\dagger 1} Z_1^\dagger W^{\dagger 2}
+
Z^2 W_1 Z^1 Z_2^\dagger W^{\dagger 1} Z_1^\dagger +
W_2 Z^2 W_1 W^{\dagger 2} Z_2^\dagger W^{\dagger 1}
\right] \\
&- \xi \;
\left[
\zeta^1 \omega_2 Z^2 W_1 +
\omega_2 \zeta^2 W_1 Z^1 + 
\zeta^2 \omega_1 Z^1 W_2 -
\omega^{\dagger 1} \zeta_1^\dagger W^{\dagger 2} Z_2^\dagger 
\right] \\
&- \xi \;
\left[
\zeta_1^\dagger \omega^{\dagger 2} Z_2^\dagger W^{\dagger 1} +
\omega^{\dagger 2} \zeta_2^\dagger W^{\dagger 1} Z_1^\dagger +
\zeta_2^\dagger \omega^{\dagger 1} Z_1^\dagger W^{\dagger 2} -
\omega_1 \zeta^1 W_2 Z^2 
\right] \\
&- \xi \;
\left.
\left[
\zeta^1 W_2 \zeta^2 W_1 +
\omega_2 Z^2 \omega_1 Z^1 +
\zeta_1^\dagger W^{\dagger 2} \zeta_2^\dagger W^{\dagger 1} +
\omega^{\dagger 2} Z_2^\dagger \omega^{\dagger 1} Z_1^\dagger 
\right] ~
\right\rbrace ~.
\end{split}
\end{equation}
Furthermore, we can follow \cite{Benna:2008zy} and compose the matter fields into a $\SU{4}_\mathrm{R}$ fundamental representation (although it is broken to $\U{1}_\mathrm{R}$),
\begin{equation}
Y^M
=
\begin{pmatrix}
Z^1 \\
Z^2 \\
W^{\dagger 1} \\
W^{\dagger 2}
\end{pmatrix} , \hspace{0.5cm}
Y_M^\dagger
=
\begin{pmatrix}
Z_1^\dagger \\
Z_2^\dagger \\
W_1 \\
W_2
\end{pmatrix} , \hspace{0.5cm}
\Psi_M
=
\e^{-\frac{\I \pi}{4}}
\begin{pmatrix}
- \zeta^2 \\
\zeta^1 \\
\I\, \omega^{\dagger 2} \\
- \I\, \omega^{\dagger 1}
\end{pmatrix} , \hspace{0.5cm}
\Psi^{\dagger M}
=
\e^{\frac{\I \pi}{4}}
\begin{pmatrix}
- \zeta_2^\dagger \\
\zeta_1^\dagger \\
- \I\, \omega _2 \\
\I\, \omega_1
\end{pmatrix}.
\end{equation}
The superfishnet action in terms of the fields $Y^M$, $Y_M^\dagger$, $\Psi_M$ and $\Psi^{\dagger M}$ reads
\begin{equation}
\begin{split}
& S_\mathrm{SFN}
=
\mathrm{N}
\int \dd^3 x ~
\mathrm{tr}
\left\lbrace
Y_M^\dagger \square Y^M
+ \I\, \Psi^{\dagger M \alpha} \gamma^\mu_{\alpha \beta} \partial_\mu \Psi_M^{\beta}
\right. \\
&+ \xi^2
\left[
Y^1 Y_4^\dagger Y^2 Y_1^\dagger Y^4 Y_2^\dagger +
Y^1 Y_4^\dagger Y^3 Y_1^\dagger Y^4 Y_3^\dagger
+
Y^2 Y_3^\dagger Y^1 Y_2^\dagger Y^3 Y_1^\dagger +
Y^2 Y_3^\dagger Y^4 Y_2^\dagger Y^3 Y_4^\dagger
\right] \\
&- \I \xi \;
\left[
\Psi_2 \Psi^{\dagger 3} Y^2 Y_3^\dagger -
\Psi^{\dagger 3} \Psi_1 Y_3^\dagger Y^1 +
\Psi_1 \Psi^{\dagger 4} Y^1 Y_4^\dagger -
\Psi_4 \Psi^{\dagger 2} Y^4 Y_2^\dagger 
\right] \\
&+ \I \xi \;
\left[
\Psi^{\dagger 2} \Psi_3 Y_2^\dagger Y^3 -
\Psi_3 \Psi^{\dagger 1} Y^3 Y_1^\dagger  +
\Psi^{\dagger 1} \Psi_4 Y_1^\dagger Y^4 -
\Psi^{\dagger 4} \Psi_2 Y_4^\dagger Y^2
\right] \\
&+ \I \xi \;
\left.
\left[
\Psi_2 Y_4^\dagger \Psi_1 Y_3^\dagger +
\Psi^{\dagger 3} Y^2 \Psi^{\dagger 4} Y^1 -
\Psi^{\dagger 2} Y^4 \Psi^{\dagger 1} Y^3 -
\Psi_3 Y_2^\dagger \Psi_4 Y_1^\dagger 
\right] ~
\right\rbrace ~.
\end{split}
\label{eq:Action_superfishnet_Y_Psi}
\end{equation}
The action contains sextic, scalar interactions and three-dimensional versions of the Yukawa coupling that are quadratic in the fermions and quadratic in the bosons.
The elegant packing of the many interaction terms of \eqref{eq:Action_superfishnet_Y_Psi} into the superfield action \eqref{eq:Action_superfishnet_Z_W} motivates us to study the superfishnet theory in a supersymmetry-covariant way, namely in the diagrammatics of Feynman supergraphs.

Another $\beta$-deformation was obtained in \cite{Caetano:2016ydc}, which coincides with \eqref{eq:Action_superfishnet_Y_Psi} up to the $\SU{4}_\mathrm{R}$-indices.
The difference is due to different bases of the $\mathfrak{su}(4)_\mathrm{R}$ Cartan-torus, c.\,f.\ table \ref{tab:Rcharges} and table 1 in \cite{Caetano:2016ydc}.
The $\beta$-deformations here and in \cite{Caetano:2016ydc} were taken into different directions in the $\SU{2} \otimes \SU{2}$ group.

\section{The Superfishnet theory}
\label{sec:SuperfishnetTheory}
We derived the superspace action of the superfishnet theory  in section \ref{sec:DoubleScaledBetaDeformationOfABJM} and found it to be \eqref{eq:Action_superfishnet_Z_W}.
In this section we will adopt a more suitable notation, present the generalized superfishnet theory, and derive the supersymmetric Feynman rules and auxiliary supergraph relation for this theory.

In order to keep track of the different fields more efficiently, we redefine the chiral superfields as $(\mathcal{Z}^1 , \mathcal{W}_2 , \mathcal{Z}^2 , \mathcal{W}_1) = \Phi_i$ and the anti-chiral ones as $(\bar{\mathcal{Z}}_1 , \bar{\mathcal{W}}^2 , \bar{\mathcal{Z}}_2 , \bar{\mathcal{W}}^1) = \Phi_i^\dagger$.
We will refer to the index $i$ as flavor.
Additionally, we introduce Gra\ss mann-squares to extend the superspace integral of the superpotential over the full chiral and anti-chiral superspace.
Finally, the superfishnet theory takes the form
\begin{equation}
\begin{split}
S_{\mathrm{SFN}}
~=~
&\mathrm{N}
\int \dd^3 x \; \dd^2 \theta \, \dd^2 \bar{\theta} ~\;
\mathrm{tr}
\left[
- \sum_{i=1}^4 \Phi_i^\dagger \Phi_i^{\phantom{\dagger}}
+ \xi \cdot \bar{\theta}^2 \, \Phi_1 \Phi_2 \Phi_3 \Phi_4
+ \xi \cdot \theta^2 \, \Phi_1^\dagger \Phi_2^\dagger \Phi_3^\dagger \Phi_4^\dagger
\right] ~.
\end{split}
\label{eq:Action_superfishnet_Phi}
\end{equation}
Despite the decoupling of the gauge degrees of freedom, the superfields are still matrices in the bi-fundamental representation of the gauge group $\U{\mathrm{N}} \otimes \U{\mathrm{N}}$.
Due to the prefactor $\mathrm{N}$ in the superfishnet action \eqref{eq:Action_superfishnet_Phi}, the supergraphs admit a genus expansion, of which we study the toroidal order that dominates in the limit $\mathrm{N} \rightarrow \infty$.

In close analogy to \cite{Kade:2024ucz}, we propose a non-local generalization \cite{Kimura:2016irk} of the superfishnet theory, where the K\"{a}hler potential is modified by fractional derivatives,
\begin{equation}
\begin{split}
S_{\mathrm{SFN},\boldsymbol{\delta}}
= \;
&\mathrm{N}
\int \dd^3 x \; \dd^2 \theta \, \dd^2 \bar{\theta} ~\;
\mathrm{tr}
\left[
- \sum_{i=1}^4 \Phi_i^\dagger \square^{\delta_i} \Phi_i^{\phantom{\dagger}}
+ \xi \cdot \bar{\theta}^2 \, \Phi_1 \Phi_2 \Phi_3 \Phi_4
+ \xi \cdot \theta^2 \, \Phi_1^\dagger \Phi_2^\dagger \Phi_3^\dagger \Phi_4^\dagger
\right] ~.
\end{split}
\label{eq:Action_genSuperfishnet_Phi}
\end{equation}
Dimensional analysis shows that the mass dimensions of the superfields get deformed to $\left[\Phi_i\right] = [\Phi^\dagger_i ] = \frac{1}{2} - \delta_i$.
Demanding a marginal interaction implies the constraint $\sum_{i=1}^4 \delta_i = 0$ for the four deformation parameters $\delta_i$. 
The form of the action \eqref{eq:Action_genSuperfishnet_Phi} is strikingly similar to the so-called checkerboard theory \cite{Alfimov:2023vev}.
Instead of bosonic fields and spacetime integration, the generalized superfishnet action contains superfields as degrees of freedom and integration over superspace.
Lastly, we note that the generalized superfishnet reduces to the ordinary superfishnet theory by setting all deformation parameters to zero, $\delta_i = 0$.

\subsection{The generalized superpropagator and Feynman super-rules}
The generalized superpropagator in three-dimensional $\mathcal{N} = 2$ superspace can be deduced from the generalized propagators of the component fields, \eqref{eq:ComponentExpansion_ChiralSuperfields} and \eqref{eq:ComponentExpansion_AntiChiralSuperfields}, and their kinetic operators in the spacetime action \eqref{eq:Action_genSuperfishnet_Phi}.
The derivation follows very closely \cite{Kade:2024ucz}, which we adapt for the three-dimensional case at hand.
We find the generalized superpropagator
\begin{equation}
\begin{split}
\left\langle \Phi_i(z_1) \Phi_j^\dagger(z_2) \right\rangle
=
\frac{\delta_{ij}}{\mathrm{N}} \,
c_{\delta_i}\cdot
\e^{\I \left[ 
\theta_1 \gamma^\mu \bar{\theta}_1 +
\theta_2 \gamma^\mu \bar{\theta}_2 -
2 \theta_1 \gamma^\mu \bar{\theta}_2 \right]\partial_{1,\mu}}
\frac{1}{\left[ x_{12}^2 \right]^{\frac{1}{2} - \delta_i}}
=
\frac{\delta_{ij}}{\mathrm{N}} 
\frac{c_{\delta_i}}{\left[x_{1\bar{2}}^2 \right]^{\frac{1}{2} - \delta_i}} ~.
\label{eq:gen_superPropagator}
\end{split}
\end{equation}
We denote supercoordinates as $z_n = (x_n, \theta_n, \bar{\theta}_n)$ and use the abbreviations $c_{\delta_i} = (-1)^{\delta_i} 2^{- \frac{1}{2} -2\delta_i} \frac{\Gamma (\frac{1}{2} - \delta_i)}{\Gamma (1 + \delta_i)}$, $x^\mu_{12}= x^\mu_1 - x^\mu_2$ and the superconformal covariant interval 
$x_{1\bar{2}}^\mu := x_{12}^\mu+\I \left[ 
\theta_1 \gamma^\mu \bar{\theta}_1 +
\theta_2 \gamma^\mu \bar{\theta}_2 -
2 \theta_1 \gamma^\mu \bar{\theta}_2 \right]$.
Note that in the undeformed superfishnet theory, we will encounter factors $c_0 = \sqrt{\frac{\pi}{2}}$.
Even though we are in Minkowski spacetime, we did not explicitly include the $\I\varepsilon$ in the denominators of \eqref{eq:gen_superPropagator} for conciseness of notation. 

For practical reasons, it is convenient to focus on the supercoordinate-dependent part of the superpropagator and reinstate the factors $\frac{c_{\delta_i}}{\mathrm{N}}$ in the end. 
Accordingly, we consider the weight function
\begin{equation}
\frac{1}{\left[x_{1\bar{2}}^2 \right]^{u}}
=
\adjincludegraphics[valign=c,scale=1]{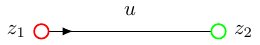} 
\label{eq:FeynmanRules_superWeight_graphical}
\end{equation}
in the following, which is the supersymmetric generalization of a QFT analog of a lattice model weight, see eq.\ (3.1) in \cite{Kade:2023xet}, and where $u\in \mathbbm{C}$ is a spectral parameter hinting at the model's integrability.
Whenever the spectral parameter of a superpropagator is not specified in superspace diagrams below, we understand it to be set to the three-dimensional default value $u=\frac{1}{2}$.
Partially, we will affix the flavor index $i$ to the superpropagator; we hope that the difference between a spectral parameter and a flavor index will be clear from the context.
As expected, the superpropagator is invariant under \eqref{eq:U1_intakt_transformation}, meaning it has zero R-charge. 
Besides the propagator, we will encounter two other formal superspace two-point functions, 
\begin{equation}
\frac{\theta_{12}^2}{\left[x_{12}^2 \right]^{u}} =
\adjincludegraphics[valign=c,scale=0.7]{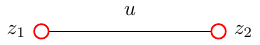} ~,
\hspace{1cm}
\frac{\bar{\theta}_{12}^2}{\left[x_{12}^2 \right]^{u}} =
\adjincludegraphics[valign=c,scale=0.7]{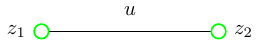} ~,
\label{eq:FeynmanRules_AuxTwoPointFctns}
\end{equation}
which have R-charge $-2$ and $2$, and represent chiral and anti-chiral delta functions on the fermionic subspace, respectively, and we introduced the abbreviation $\theta_{12} = \theta_1 - \theta_2$.
The dependence on the bosonic coordinates is that of a bosonic generalized propagator.
Therefore, there is the possibility that the spectral parameter approaches $\frac{3}{2}$.
Assuming the proper normalization, by the representation of the three-dimensional bosonic delta function \cite{Kade:2023xet}
\begin{equation}
\delta^{(3)}\left( x_{12} \right) 
~=~
\lim_{\varepsilon \rightarrow 0} ~ \pi^{-\frac{3}{2}} 
\frac{\Gamma (\frac{3}{2} - \varepsilon)}{\Gamma (\varepsilon)}
\cdot \frac{1}{\left[ x_{12}^2 \right]^{\frac{3}{2} - \varepsilon}} ~,
\label{eq:DeltaDefi}
\end{equation}
the two-point functions \eqref{eq:FeynmanRules_AuxTwoPointFctns} respectively can turn into chiral and anti-chiral superspace delta functions.
Graphically, we denote them with a dashed line, 
\begin{equation}
\delta^{(2)}\left(\theta_{12}\right) \delta^{(3)}\left( x_{12} \right) =
\adjincludegraphics[valign=c,scale=0.9]{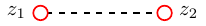} ~,
\hspace{0.2cm}
\delta^{(2)}\left(\bar{\theta}_{12}\right) \delta^{(3)}\left( x_{12} \right) =
\adjincludegraphics[valign=c,scale=0.9]{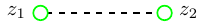} ~.
\label{eq:SuperspaceDeltaDefi}
\end{equation}

We read off the vertex Feynman rules from the superfishnet action \eqref{eq:Action_superfishnet_Phi} or the generalization \eqref{eq:Action_genSuperfishnet_Phi} alike.
After Wick-rotating, they are
\begin{equation}
\adjincludegraphics[valign=c,scale=0.6]{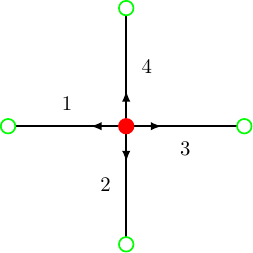}
\sim
\mathrm{N}\, \xi \cdot \I \int \dd^3x\; \dd^2\theta\,\dd^2\bar{\theta} \;\delta^{(2)}(\bar{\theta}) ~, ~
\adjincludegraphics[valign=c,scale=0.6]{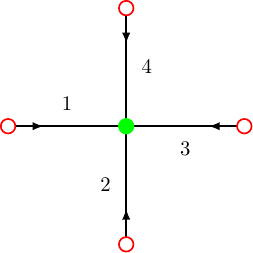}
\sim
\mathrm{N}\, \xi \cdot \I \int \dd^3x\; \dd^2\theta\,\dd^2\bar{\theta} \;\delta^{(2)}(\theta) ~.
\label{eq:FeynmanRules_vertices}
\end{equation}
Firstly, note that the labels in \eqref{eq:FeynmanRules_vertices} are the flavors of the superfields, which meet at the respective vertex.
They should not be confused with the spectral parameter of the superpropagator.
Due to the non-unitarity of the superfishnet theory \eqref{eq:Action_superfishnet_Phi}, the two vertices cannot be mapped into each other by hermitian conjugation.
Respecting the order of the flavor in each vertex reduces the number of allowed supergraphs and leads to the regular fishnet pattern.
Secondly, similar to the superpropagators, we will neglect the factor $\mathrm{N}\xi$ in our calculations for now.
In the end, we reinstate them for the diagram under consideration.
Thirdly, here and in the following, we will denote internal, integrated, chiral (anti-chiral) vertices by a filled red (green) dot corresponding to the left (right) vertex in \eqref{eq:FeynmanRules_vertices}.
Note that the Gra\ss mann delta functions in \eqref{eq:FeynmanRules_vertices} annihilate the part of the fermionic integration whose chirality is opposite to the one of the vertex at hand.
Hence, propagators always connect the chiral and anti-chiral subspaces of superspace.
When we expect an external point in a super Feynman diagram to be integrated by a chiral (anti-chiral) vertex in a later step according to the Feynman vertex rules \eqref{eq:FeynmanRules_vertices}, we denote this by an un-filled red (green) circle.
A black, un-filled circle indicates an external point in full superspace.

\subsection{Super chain relations}
\label{subsec:SuperChainRelations}
A handy tool in calculating supergraphs are the super chain relations.
They allow the reduction of the super-convolution of two two-point functions to a single one.
In \cite{Kade:2024ucz}, they are shown by direct superspace-integration; however, now the bosonic integral is three-dimensional.
At first, we consider the convolution of two generalized superpropagators \eqref{eq:FeynmanRules_superWeight_graphical}.
We find their anti-chiral chain rule\footnote{Note that we can lift the restriction to chiral external superspace points by action with the operators $\e^{\I \theta_1 \gamma^\mu \bar{\theta}_1 \partial_{1,\mu}}$ and $\e^{\I \theta_2 \gamma^\mu \bar{\theta}_2 \partial_{2,\mu}}$. Accordingly, the chiral chain rule can be lifted by similar operators.}
\begin{subequations}
\begin{align}
\left[
\I \int \dd^3x_0\, \dd^2\bar{\theta}_0
\frac{1}{\left[x_{1\bar{0}}^2 \right]^{u_1}}
\frac{1}{\left[x_{2\bar{0}}^2 \right]^{u_2}}
\right]_{\substack{\theta_0 = 0 \\ \bar{\theta}_{1,2}=0}}
=
4\I\, r ( 2 - u_1 - u_2 , u_1 , u_2 )\;
\frac{\theta_{12}^2}{\left[x_{12}^2 \right]^{u_1+u_2 - \frac{1}{2}}} ~, \\
\adjincludegraphics[valign=c,scale=1]{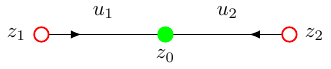}
= 
4\I\, r ( 2 - u_1 - u_2 , u_1 , u_2 )\;
\adjincludegraphics[valign=c,scale=1]{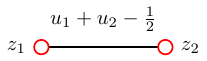} ~,
\end{align}
\label{eq:ChainRelAntiChiral}%
\end{subequations}
and its chiral counterpart
\begin{subequations}
\begin{align}
\left[
\I \int \dd^3x_0\, \dd^2\theta_0
\frac{1}{\left[x_{0\bar{1}}^2 \right]^{u_1}}
\frac{1}{\left[x_{0\bar{2}}^2 \right]^{u_2}}
\right]_{\substack{\bar{\theta}_0 = 0 \\ \theta_{1,2}=0}}
=
4\I\, r ( 2 - u_1 - u_2 , u_1 , u_2 )\;
\frac{\bar{\theta}_{12}^2}{\left[x_{12}^2 \right]^{u_1+u_2 - \frac{1}{2}}} ~, \\
\adjincludegraphics[valign=c,scale=1]{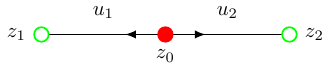}
= 
4\I\, r ( 2 - u_1 - u_2 , u_1 , u_2 )\;
\adjincludegraphics[valign=c,scale=1]{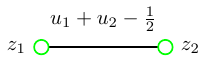} ~.
\end{align}
\label{eq:ChainRelChiral}%
\end{subequations}
We introduce the abbreviations for factors of $\Gamma$-functions 
\begin{equation}
r(u_1,u_2,u_3):= \pi^{\frac{3}{2}}\; a (u_1)\, a (u_2)\, a (u_3)
~~\mathrm{with}~~
a (u) := \frac{\Gamma (\frac{3}{2} - u )}{\Gamma ( u )} ~.
\label{eq:RfactorAfactor}
\end{equation}
We observe that in comparison with the four-dimensional $\mathcal{N}=1$ chain rules \cite{Kade:2024ucz}, the argument in the $r$ factor in \eqref{eq:ChainRelAntiChiral} and \eqref{eq:ChainRelChiral} is reduced to $2$.
The reason is that the scaling dimension of the integration measure is reduced since the integrals are performed in three bosonic dimensions.

Next, there exists a critical limiting case of \eqref{eq:ChainRelAntiChiral} and \eqref{eq:ChainRelChiral}, which is when $u_1$ and $u_2$ are chosen to add up to $2$.
Then, the r.\ h.\ s.\ also becomes a delta function on the bosonic space.
The resulting relations are 
\begin{subequations}
\begin{align}
\lim_{\varepsilon \rightarrow 0}
\adjincludegraphics[valign=c,scale=1]{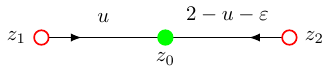}
&= 
4\I\,\pi^{3} \cdot a ( u )\, a ( 2 - u )
\adjincludegraphics[valign=c,scale=1]{pictures/basics/deltadist/chiral_delta/DeltaDist.pdf} ~,
\label{eq:ResolutionOfUnity_antichiral}\\
\lim_{\varepsilon \rightarrow 0}
\adjincludegraphics[valign=c,scale=1]{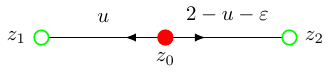}
&= 
4\I\,\pi^{3} \cdot a( u )\, a ( 2 - u )
\adjincludegraphics[valign=c,scale=1]{pictures/basics/deltadist/antichiral_delta/DeltaDist.pdf} ~. \label{eq:ResolutionOfUnity_chiral}
\end{align}
\label{eq:ResolutionOfUnity}%
\end{subequations}

We now consider convolutions of a superpropagator \eqref{eq:FeynmanRules_superWeight_graphical} with an R-charged two-point function \eqref{eq:FeynmanRules_AuxTwoPointFctns}.
Starting with the anti-chiral version, we find 
\begin{subequations}
\begin{align}
&\left[
\I \int \dd^3x_0\, \dd^2\bar{\theta}_0
\frac{1}{\left[x_{1\bar{0}}^2 \right]^{u_1}}
\frac{\bar{\theta}_{02}^2}{\left[x_{20}^2 \right]^{u_2}}
\right]_{\substack{\theta_{0,2} = 0 \\ \bar{\theta}_{1}=0}}
=
\I\, r ( 3 - u_1 - u_2 , u_1 , u_2 )\;
\frac{1}{\left[x_{1\bar{2}}^2 \right]^{u_1+u_2 - \frac{3}{2}}} ~, \\
&\adjincludegraphics[valign=c,scale=1]{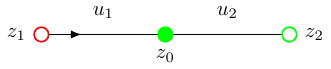}
= 
\I\, r ( 3 - u_1 - u_2 , u_1 , u_2 )\;
\adjincludegraphics[valign=c,scale=1]{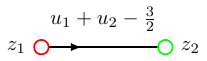} ~,
\end{align}
\label{eq:ChainRelAntiChiral_Aux}%
\end{subequations}
and its chiral counterpart
\begin{subequations}
\begin{align}
&\left[
\I \int \dd^3x_0\, \dd^2\theta_0
\frac{1}{\left[x_{0\bar{1}}^2 \right]^{u_1}}
\frac{\theta_{02}^2}{\left[x_{20}^2 \right]^{u_2}}
\right]_{\substack{\bar{\theta}_{0,2} = 0 \\ \theta_{1}=0}}
=
\I\, r ( 3 - u_1 - u_2 , u_1 , u_2 )\;
\frac{1}{\left[x_{2\bar{1}}^2 \right]^{u_1+u_2 - \frac{3}{2}}} ~, \\
& \adjincludegraphics[valign=c,scale=1]{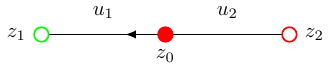}
= 
\I\, r ( 3 - u_1 - u_2 , u_1 , u_2 )\;
\adjincludegraphics[valign=c,scale=1]{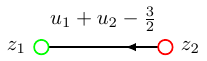} ~.
\end{align}
\label{eq:ChainRelChiral_Aux}%
\end{subequations}

\subsection{Super x-unity relation}
The super x-unity relation is the pivotal auxiliary relation for calculating the critical coupling in section \ref{subsec:InversionRelations}.
It allows us to find the inverse of the graph-building row-matrix of the superfishnet vacuum graphs and to apply the method of inversion relations.
The derivation of the super x-unity relation for the superfishnet theory is very similar to the one of the super brick wall theory \cite{Kade:2024ucz}.
However, we will not rely on the superconformal star-integral due to Osborn \cite{Osborn:1998qu,Dolan:2000uw}.

Let us make two observations before deriving the super x-unity relation. 
\begin{itemize}
\item 
Consider a three-spiked star integral, where the weights add up to $2$, and where we take one superpropagator weight to zero.
Then the propagator disappears in the limit, and we obtain the integral relation \eqref{eq:ResolutionOfUnity_antichiral}.
This yields the relation
\begin{equation}
\lim_{\varepsilon\rightarrow 0}
\adjincludegraphics[valign=c,scale=1]{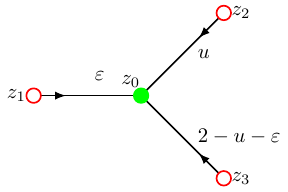} 
=
4\I\,\pi^3
a(u)\, a(2 - u) ~\cdot
\adjincludegraphics[valign=c,scale=1]{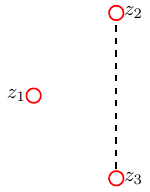} ~.
\label{eq:3ptFctnWithOneVanishingParameter}
\end{equation}

\item
We obtain another helpful relation when integrating one external point of a three-spiked star integral over the chiral subspace of superspace, in the case where the propagator weights add up to $2$: 
\begin{equation}
\adjincludegraphics[valign=c,scale=1]{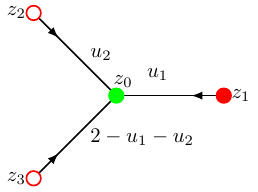} 
=
4\pi^3
a(u_1)\, a(2 - u_1)
\adjincludegraphics[valign=c,scale=1]{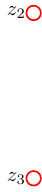} 
\label{eq:3ptFctnOnePointIntegrated}
\end{equation}
The derivation of the relation \eqref{eq:3ptFctnOnePointIntegrated} is shown in detail in appendix \ref{subsec:auxRelForXUnity}.
We observe that the whole expression reduces to a factor independent of $u_2$.
\end{itemize}
With the help of the relations \eqref{eq:ResolutionOfUnity_chiral}, \eqref{eq:3ptFctnWithOneVanishingParameter} and \eqref{eq:3ptFctnOnePointIntegrated} we can follow the steps of appendix B in \cite{Kade:2024ucz} to derive the super x-unity relations for the three-dimensional case at hand.
They are
\begin{subequations}
\begin{align}
\adjincludegraphics[valign=c,scale=1]{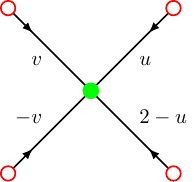} 
=
4\,\pi^3
a(u)\, a(2-u) \hspace{0.5cm}
\adjincludegraphics[valign=c,scale=1]{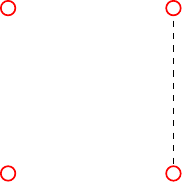} ~,\\
\adjincludegraphics[valign=c,scale=1]{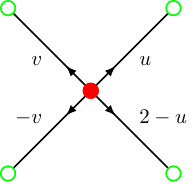} 
=
4\,\pi^3
a(u)\, a(2-u) \hspace{0.5cm}
\adjincludegraphics[valign=c,scale=1]{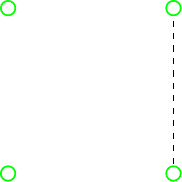} 
 ~.
\end{align}\label{eq:SuperXUnity}%
\end{subequations}

\section{Vacuum graphs in the thermodynamic limit}
\begin{figure}[h]
\includegraphics[scale=0.5]{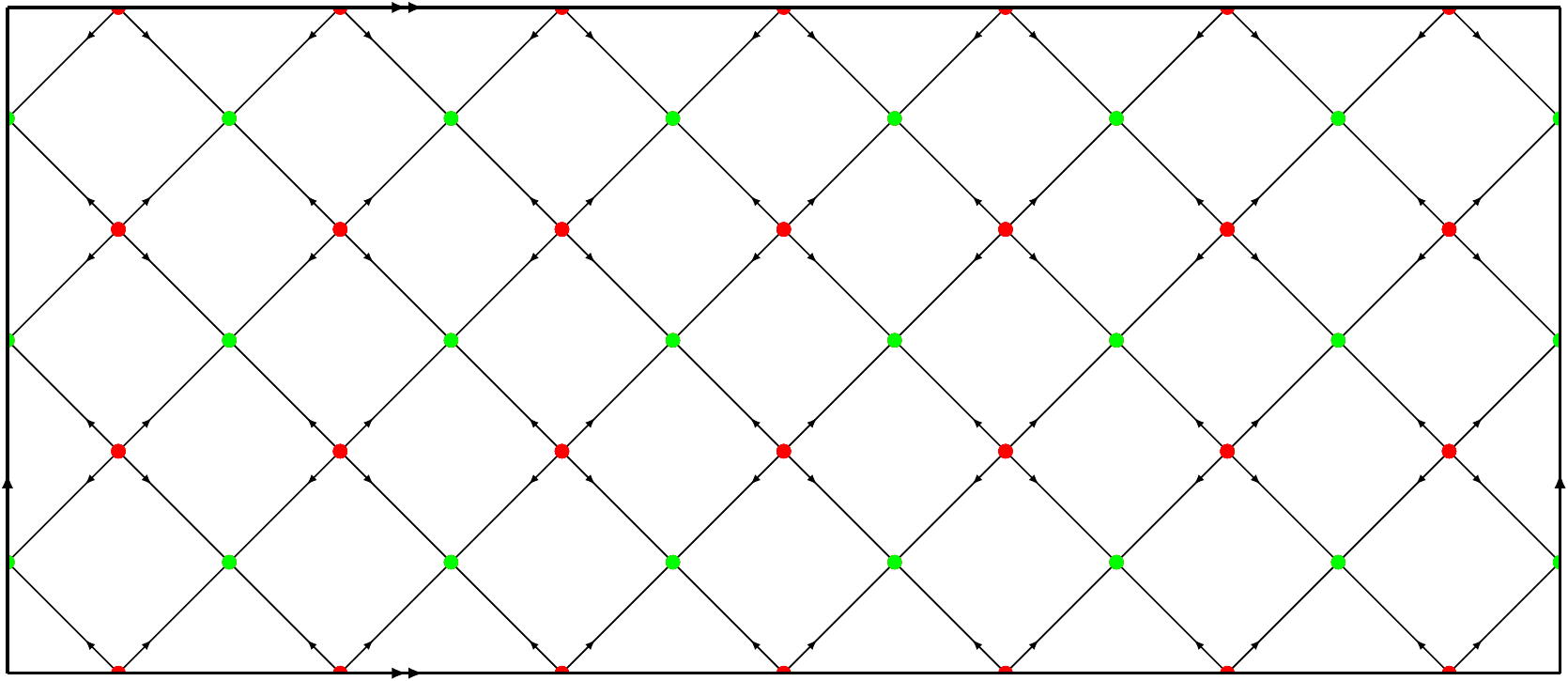} \centering
\caption{The toroidal super vacuum graph $Z_{3,7}$ exemplifies a contribution to the superfishnet's free energy.
% It contains three rows, each consisting of four graph building kernels \eqref{eq:GraphBuilderBW}. 
The faces of the graph are squares, and the vacuum supergraphs resemble a fishing net, which gives the theory its name. 
In the limit $\mathrm{N}\rightarrow\infty$, the leading order results in toroidal diagrams. This is implemented into the figure: 
%the thick lines bounding the figure should be identified according to the number of arrows on them: 
top and bottom lines are identified, and so are the left and right boundary lines.}
\label{fig:betaDef_vacuumgraph}
\end{figure}
An example of a vacuum supergraphs of the superfishnet theory \eqref{eq:Action_superfishnet_Phi} in the planar limit is shown in fig.\ \ref{fig:betaDef_vacuumgraph}. 
The regular fishnet pattern is due to the highly constraining Feynman rules and, notably, the interaction vertices \eqref{eq:FeynmanRules_vertices} in the action \eqref{eq:Action_superfishnet_Phi}. 
Diagrammatically, the generalized superfishnet theory \eqref{eq:Action_genSuperfishnet_Phi} reproduces the same graphs as in fig.\ \ref{fig:betaDef_vacuumgraph} because the superpotential is the same as in the undeformed theory \eqref{eq:Action_superfishnet_Phi}.
However, the propagator weights differ by their exponent, and one has to multiply the propagators with the prefactor $c_{\delta_i}$.

A critical comment is in place: vacuum diagrams in field theory are generally proportional to the spacetime volume, and thus, the overall free energy is usually infrared divergent. 
Since we are only interested in the free energy density, this divergence is best fixed by leaving one of the points in the vacuum graphs unintegrated, which fixes the zero mode.
Similarly, superspace vacuum graphs are proportional to the ill-defined $\int \dd^3 x\cdot\int \dd^2 \theta\, \dd^2 \bar{\theta} = \infty \cdot 0$.
(The zero stemming from the fermionic integration is just a manifestation of the well-known statement that the vacuum energy of supersymmetric field theories is zero.)
In our supersymmetric model at hand, we proceed in the same way and leave one of the superspace points in the vacuum supergraphs un-integrated in order to obtain a well-defined density. 

\begin{figure}[h]
\includegraphics[scale=1]{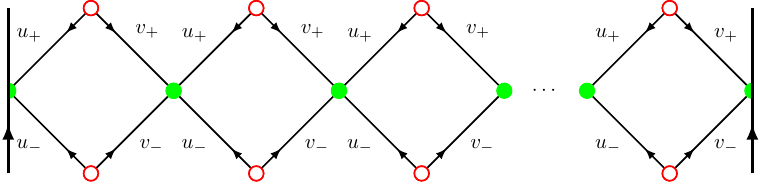}  \centering
\caption{The generalized row matrix $T_N(\mathbf{u})$ contains anti-chiral internal vertices (green, filled). Its external points are chiral, un-integrated vertices (red, un-filled). The row matrix depends on four spectral parameters, collectively denoted as $\mathbf{u}=\left(\begin{smallmatrix}u_+ & v_+ \\ u_- & v_- \end{smallmatrix}\right)$. Multiple copies of the row matrix can be stacked on top of each other to build up a generalized super Feynman graph according to \eqref{eq:gen_vacuumdiagram}. The ``matrix product'' thereby consists of $N$ integrals over the chiral part of superspace.}
\label{fig:betaDef_gen_rowmatrix}
\end{figure}

Starting from the super weight \eqref{eq:FeynmanRules_superWeight_graphical}, a superpropagator without the factor $c_{\delta_i}$, we can identify\footnote{ 
Note, that we could have chosen another row matrix kernel with the chiralities interchanged, i.\,e.\ with anti-chiral (green) external vertices and chiral (red) internal ones.
}
a generalized row-matrix $T_N(\mathbf{u})$, see fig.\ \ref{fig:betaDef_gen_rowmatrix}, building up the vacuum diagrams in fig.\ \ref{fig:betaDef_vacuumgraph} after e.\,g.\ fixing the parameters to $\mathbf{u}:=\left(\begin{smallmatrix}u_+ & v_+ \\ u_- & v_- \end{smallmatrix}\right)=\left(\begin{smallmatrix}\sfrac{1}{2} & \sfrac{1}{2} \\ \sfrac{1}{2} & \sfrac{1}{2} \end{smallmatrix}\right)$.
Formally, we may write a generalized $M\x N$ toroidal vacuum supergraph as 
\begin{equation}
Z_{MN}(\mathbf{u}) = \mathrm{tr}\left[ T_N(\mathbf{u})^M \right],
\label{eq:gen_vacuumdiagram}
\end{equation}
which graphically represents $M$ generalized row matrices of length $N$ stacked on top of each other and identified periodically by the trace. 
However, the factor $c_0$ is not included in the vacuum graph.

To reinstate the factor $c_0$, and for studying the more general supergraphs of theory \eqref{eq:Action_genSuperfishnet_Phi}, we have to consider an enhanced row matrix\footnote{Remember that $\delta_1 + \delta_2 + \delta_3 + \delta_4 = 0$, which annihilates factors with the power of $\delta_i$ in $c_{\delta_i}$.
}
\begin{equation}
\mathbbm{T}_N (\mathbf{u}) 
=
\left[
\frac{1}{4} a(\tfrac{3}{2}- u_+) a(\tfrac{3}{2}- u_-) a(\tfrac{3}{2}- v_+) a(\tfrac{3}{2}- v_-)
\right]^N
T_N(\mathbf{u}) ~.
\label{eq:gen_RowMatrix_omega}
\end{equation}
We denote the generalized $M\x N$ toroidal vacuum supergraphs built by the enhanced row matrix by
\begin{equation}
\mathbbm{Z}_{MN}(\mathbf{u}) = \mathrm{tr}\left[ \mathbbm{T}_N(\mathbf{u})^M \right] ~.
\label{eq:gen_vacuumdiagram_enhanced}
\end{equation}
To obtain the vacuum graphs of the generalized superfishnet theory, we have to tune the parameters to e.\,g.\ $\mathbf{u} \rightarrow \boldsymbol{\delta}:=\left(\begin{smallmatrix} \sfrac{1}{2} - \delta_1 & \sfrac{1}{2} - \delta_4 \\ \sfrac{1}{2} - \delta_2 & \sfrac{1}{2} - \delta_3 \end{smallmatrix}\right)$.
Conversely, to obtain ordinary superfishnet theory, we have to turn off all deformation parameters, $\delta_i = 0$.
By way of example, fig.\ \ref{fig:betaDef_vacuumgraph} shows the resulting graph for $M=3$ and $N=7$, called $\mathbbm{Z}_{3,7}\left(\begin{smallmatrix}\sfrac{1}{2} & \sfrac{1}{2} \\ \sfrac{1}{2} & \sfrac{1}{2} \end{smallmatrix}\right) = c_0^{84}Z_{3,7}\left(\begin{smallmatrix}\sfrac{1}{2} & \sfrac{1}{2} \\ \sfrac{1}{2} & \sfrac{1}{2} \end{smallmatrix}\right)$.

We aim at calculating the critical coupling $\xi_\mathrm{cr}$. It is defined as the radius of convergence of the expansion of the free energy of \eqref{eq:Action_genSuperfishnet_Phi} (including the special case \eqref{eq:Action_superfishnet_Phi}), which is 
\begin{equation}
Z_{\boldsymbol{\delta}} ~=~ 
\sum_{M,N=1}^\infty \mathbbm{Z}_{MN}\left(\boldsymbol{\delta}\right) \cdot 
\xi^{2MN} ~ .
\end{equation}
The critical coupling is then $\xi_\mathrm{cr}= \left[ \mathbbm{K}\left(\boldsymbol{\delta}\right) \right]^{-1/2}$, where we evaluate the generalized vacuum diagrams in the thermodynamic limit 
\begin{equation}
\mathbbm{K}(\mathbf{u}) ~:=~
\lim_{M,N\rightarrow \infty} \vert \mathbbm{Z}_{MN}(\mathbf{u})\vert^{\frac{1}{MN}}
\label{eq:gen_freeEnergy}
\end{equation}
at $\mathbf{u}=\boldsymbol{\delta}$ and eventually at $\delta_i \rightarrow 0$. 

We will calculate the limit \eqref{eq:gen_freeEnergy} by the method of inversion relations.
As a first step, this requires finding the inverse of the row matrix $T_N(\mathbf{u})$ and determining the limit $K (\mathbf{u}):=\lim_{M,N\rightarrow \infty} \vert Z_{MN}(\mathbf{u})\vert^{\frac{1}{MN}}$.
Subsequently, we make the connection to \eqref{eq:gen_freeEnergy} by reinstating the additional factors in \eqref{eq:gen_RowMatrix_omega} via the relation
\begin{equation}
\mathbbm{K}(\mathbf{u}) 
~=~
\left[
\frac{1}{4} a(\tfrac{3}{2}- u_+) a(\tfrac{3}{2}- u_-) a(\tfrac{3}{2}- v_+) a(\tfrac{3}{2}- v_-)
\right]
K(\mathbf{u}) ~.
\end{equation}

\subsection{Inversion relations}
\label{subsec:InversionRelations}
The super x-unity relation allows us to find four different forms of the inverse of the row matrix $T_N(\mathbf{u})$ by the same moves explained in section 5 of \cite{Kade:2023xet}. We find
\begin{equation}
T_N(\mathbf{u})
\circ
T_N(\mathbf{u}_{\mathrm{inv}})
=
F_N \cdot \mathbbm{1}_N
\label{eq:RowMatrixInversion}
\end{equation}
to hold for 
\begin{subequations}
\begin{align}
\mathbf{u}_\mathrm{inv}&=\left(\begin{smallmatrix} -u_- & 2-v_- \\ 2-u_+ & -v_+ \end{smallmatrix}\right)~~\mathrm{and}~~F_N=\left[ 16\pi^6\; a(u_+)\, a(2-u_+)\, a(v_-)\, a(2-v_-) \right]^N ~, \\
\mathbf{u}_\mathrm{inv}&=\left(\begin{smallmatrix} -u_- & 2-v_- \\ -u_+ & 2-v_+ \end{smallmatrix}\right)~~\mathrm{and}~~F_N=\left[ 16\pi^6\; a(v_+)\, a(2-v_+)\, a(v_-)\, a(2-v_-) \right]^N ~, \\
\mathbf{u}_\mathrm{inv}&=\left(\begin{smallmatrix} 2-u_- & -v_- \\ 2-u_+ & -v_+ \end{smallmatrix}\right)~~\mathrm{and}~~F_N=\left[ 16\pi^6\; a(u_+)\, a(2-u_+)\, a(u_-)\, a(2-u_-) \right]^N ~,\\
\mathbf{u}_\mathrm{inv}&=\left(\begin{smallmatrix} 2-u_- & -v_- \\ -u_+ & 2-v_+ \end{smallmatrix}\right)~~\mathrm{and}~~F_N=\left[ 16\pi^6\; a(u_-)\, a(2-u_-)\, a(v_+)\, a(2-v_+) \right]^N .
\end{align}
\end{subequations}
The maximal eigenvalue dominates in the thermodynamic limit \cite{Kade:2023xet}, and we require it to satisfy the same functional relations as the row-matrix \eqref{eq:RowMatrixInversion}.
Then, we can turn \eqref{eq:RowMatrixInversion} into four functional relations for $K(\mathbf{u})$. 
We make the ansatz $K(\mathbf{u}) = \kappa_1(u_+)\kappa_2(u_-)\kappa_3(v_+)\kappa_4(v_-)$ and find $\kappa_i(u) \equiv \kappa(u)$ for all $i$, which has to satisfy
\begin{equation}
\kappa(u)\kappa(-u) = 1
~~~
\mathrm{and}
~~~
\kappa(u)\kappa(2-u) = 4\pi^3\; a(u)\, a(2-u) 
= 4 \pi ^3 \frac{\Gamma ( \frac{3}{2}-u ) \Gamma ( u-\frac{1}{2} )}{\Gamma (u) \Gamma (2-u)} ~.
\end{equation}
Using the method explained in \cite{Kade:2023xet}, these functional relations are solved by the function
\begin{equation}
\kappa (u) 
=
2^{\frac{3 u}{2}-1} \pi ^{\frac{3 u}{2}-\frac{1}{2}} 
\frac{\Gamma \left(\frac{3}{2}-u\right) \Gamma \left(\frac{u}{2}+\frac{1}{4}\right)}{\Gamma \left(\frac{5}{4}\right)}
\prod _{k=1}^{\infty } 
\frac
{\Gamma (2 k - u + \frac{3}{2} ) \Gamma ( 2 k + u ) \Gamma (2 k - \frac{3}{2} )}
{\Gamma (2 k + u - \frac{3}{2} ) \Gamma ( 2 k - u ) \Gamma (2 k + \frac{3}{2} )} ~,
\end{equation}
with the special values $\kappa (0) = 1$ and $\kappa (\tfrac{1}{2}) = \left(\frac{\pi}{2} \right)^{1/4} \Gamma (\tfrac{1}{4})$.
Finally, we find the critical coupling
\begin{equation}
\xi_\mathrm{cr}
=
\left[ \mathbbm{K}\left(\begin{smallmatrix} \sfrac{1}{2} - \delta_1 & \sfrac{1}{2} - \delta_4 \\ \sfrac{1}{2} - \delta_2 & \sfrac{1}{2} - \delta_3 \end{smallmatrix}\right) \right]^{-1/2}
= 
\left[
\frac{1}{4} \prod_{i = 1}^4 a(1 - \delta_i) \kappa (\tfrac{1}{2} - \delta_i )
\right]^{-1/2} ~
\label{eq:CritCoup_SFN_gen}
\end{equation}
for the generalized superfishnet theory \eqref{eq:Action_genSuperfishnet_Phi} and 
\begin{equation}
\xi_\mathrm{cr}
=
\left[ \mathbbm{K}\left(\begin{smallmatrix} \sfrac{1}{2} & \sfrac{1}{2} \\ \sfrac{1}{2} & \sfrac{1}{2} \end{smallmatrix}\right) \right]^{-1/2}
= 
\frac{2}{\Gamma (\frac{1}{2})^2} \cdot
\kappa (\tfrac{1}{2})^{-2}
=
\frac{\left(\frac{2}{\pi}\right)^{3/2}}{\Gamma \left(\frac{1}{4}\right)^2} ~
\label{eq:CritCoup_DS_gen}
\end{equation}
for the superfishnet theory \eqref{eq:Action_superfishnet_Phi}.

\section{Exact all-loop anomalous dimensions from four-point correlation functions}
\label{sec:ExactAnomalousDimensionsFrom4PtCorrelationFunctions}
This section is devoted to the study of four-point functions in the undeformed $\boldsymbol{\delta} = 0$ superfishnet theory that admit a perturbative expansion in regular, fishnet Feynman supergraphs in the large-$\mathrm{N}$ limit.
We follow the strategy presented in \cite{Grabner:2017pgm,Kazakov:2018qbr,Gromov:2018hut,Kazakov:2018gcy} to extract exact scaling dimensions out of the four-point correlators:
The correlators under investigation can be built up by repetitive application of a specific supergraph-building operator.
The operator can be resummed in a geometric series and diagonalized on a set of eigenfunctions.
The resummed series has poles, which correspond to exchanged operators.
We identify the position of the poles and may access the anomalous dimension of specific operators.
They are the ones where we can construct the corresponding eigenfunctions of the graph-building operator.

One class of three-dimensional $\mathcal{N}=2$ eigenfunctions were derived in \cite{Chang:2021fmd}.
For another class, however, we are unaware of the eigenfunctions' precise form, yet we can construct some of them in a particular limit. 
In both cases, the limit is enough to compute the eigenvalue of the correlator's graph-builder and to calculate the anomalous dimension of some operators; this is presented in appendix \ref{subsec:DiagonalizationOfTheSuperGraphBuildingOperators}.
In the so-called zero-magnon case in section \ref{subsec:ZeroMagnonCase}, we give a more detailed outline of the procedure to be concise in the subsequent two-magnon case in section \ref{subsec:TwoMagnonCase}.
To support our method of determining the anomalous dimensions, we show in appendix \ref{subsubsec:ZeroMagnonSBW} that in the case of the super brick wall theory, we reproduce results from the literature \cite{Kazakov:2018gcy}.
The calculation is based on the findings of \cite{Fitzpatrick:2014oza,Khandker:2014mpa,Li:2016chh} for four-dimensional $\mathcal{N}=1$ superconformal symmetry.

\subsection{Zero-magnon case}
\label{subsec:ZeroMagnonCase}
The zero-magnon case is the four-point correlator $
\left\langle \mathrm{tr} \left[\Phi_1 (z_1)\Phi_3^\dagger (z_2)\right] \mathrm{tr}\left[\Phi_1^\dagger (z_3)\Phi_3 (z_4)\right]\right\rangle $ in the large-$\mathrm{N}$ limit.
Equivalently, we could also have chosen the correlator with flavors $(\mathrm{1},\mathrm{3}) \leftrightarrow (\mathrm{2},\mathrm{4})$ interchanged, where the calculation is essentially unchanged.
The Feynman supergraphs are of toroidal order in the genus expansion, i.\,e.\ they scale as $\mathrm{N}^0$.
Due to the restricting flavor-ordering in the vertices \eqref{eq:FeynmanRules_vertices}, the diagrams have a very regular, ladder-like structure,
\begin{equation}
\begin{split}
\left\langle 
\mathrm{tr}
\left[
\Phi_1 (z_1)
\Phi_3^\dagger (z_2)
\right]
\mathrm{tr}
\left[
\Phi_1^\dagger (z_3)
\Phi_3 (z_4)
\right]
\right\rangle 
=
\adjincludegraphics[valign=c,scale=1]{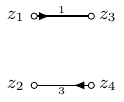} 
+
\xi^2
\adjincludegraphics[valign=c,scale=1]{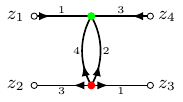} \\
+
\xi^4
\adjincludegraphics[valign=c,scale=1]{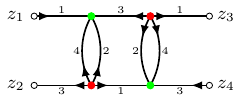} 
+
\xi^6
\adjincludegraphics[valign=c,scale=1]{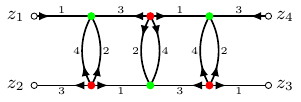} 
+
%\adjincludegraphics[valign=c,scale=1]{pictures/2ptfctn/zeromagnon/diags/4ptfunction_4.pdf} 
%+ 
\cdots ~.
\end{split}
\label{eq:0MagnonPertExpansion}
\end{equation}
Note that in the odd terms, the superpoints $z_3$ and $z_4$ are interchanged with respect to the even terms.
The diagrammatics is similar to bi-scalar fishnet four-point functions, however, here we are considering supergraphs.
The description ``zero-magnon'' refers to the closed, vertical subgraphs made up of the flavor-two and flavor-four superpropagators, which run in circles and do not contract to the operators at the external points of the diagrams.
In the bi-scalar fishnet theory, they are also called mirror-magnons \cite{Basso:2018agi,Basso:2019xay}.

We can identify a graph-building operator $\mathbb{H}$ that, together with its conjugate $\bar{\mathbbm{H}}$ and the permutation operator of two points $\mathbbm{P}$, build up the graphs of \eqref{eq:0MagnonPertExpansion}.
They are
\begin{equation}
\begin{aligned}[c]
\mathbbm{H}
=
\adjincludegraphics[valign=c,scale=1]{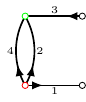} ~,
\end{aligned}
\qquad\qquad
\begin{aligned}[c]
\bar{\mathbbm{H}}
=
\adjincludegraphics[valign=c,scale=1]{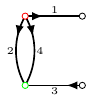} ~,
\end{aligned}
\begin{aligned}[c]
\qquad \qquad
\mathbbm{P}
=
\adjincludegraphics[valign=c,scale=1]{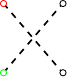} ~,
\end{aligned}
\label{eq:0MagnonGraphBuilder}
\end{equation}
and we can use them to rewrite the correlation function \eqref{eq:0MagnonPertExpansion} formally as
\begin{equation}
\begin{split}
&\left\langle 
\mathrm{tr}
\left[
\Phi_1 (z_1)
\Phi_3^\dagger (z_2)
\right]
\mathrm{tr}
\left[
\Phi_1^\dagger (z_3)
\Phi_3 (z_4)
\right]
\right\rangle \\
&=
\adjincludegraphics[valign=c,scale=1]{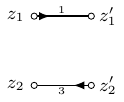} 
\circ
\left[
1
+
\xi^2
\mathbbm{H} \circ \mathbbm{P}
+
\xi^4
\mathbbm{H} \circ \bar{\mathbbm{H}}
+
\xi^6
\mathbbm{H} \circ \bar{\mathbbm{H}} \circ \mathbbm{H} \circ \mathbbm{P}
+
%\adjincludegraphics[valign=c,scale=1]{pictures/2ptfctn/zeromagnon/diags/4ptfunction_4.pdf} 
%+ 
\cdots
\right] \\
&=
\adjincludegraphics[valign=c,scale=1]{pictures/2ptfctn/zeromagnon/diags/4ptfunction_deco0_prime.pdf} 
\circ
\left[
\sum_{n = 0}^\infty 
(\xi^4 \mathbbm{H} \circ \bar{\mathbbm{H}} )^n
\right]
\left[
1 + \xi^2 \mathbbm{H} \circ \mathbbm{P}
\right] 
 ~.
\end{split}
\label{eq:0MagnonPertExpansion_2}
\end{equation}
The symbol $\circ$ denotes the super-convolution of chiral or anti-chiral superspace points. 
Its precise meaning can be inferred from the supergraph expansion \eqref{eq:0MagnonPertExpansion}.
We observe that only the combinations $\mathbbm{H} \circ \bar{\mathbbm{H}}$ and $\mathbbm{H} \circ \mathbbm{P}$ of the graph-builders \eqref{eq:0MagnonGraphBuilder} appear in \eqref{eq:0MagnonPertExpansion_2}.
Therefore, we would like to diagonalize them over a complete set of eigenfunctions.

In the bosonic case, e.\,g.\ the bi-scalar fishnet theory, the complete set of eigenfunctions are so-called conformal triangles \cite{Dobrev:1977qv,Polyakov:1970xd,Fradkin:1978pp}.
They are three-point functions that furnish a non-compact representation of the conformal group.
Taking two conformal triangles and integrating them over a common space point, and integrating/summing over their Cartan labels gives a completeness relation.
We assume the existence of a superspace generalization of this construction, which means that there exists a superconformal three-point function
\begin{equation}
\Omega_{\Delta,S,R} {\scriptstyle (z_1, z_2 ; z_0)}
=
\left\langle
\mathrm{tr}
\left[ 
\mathcal{O}_1  (z_1)
\mathcal{O}_2  (z_2)
\right]
\mathcal{O}_{\Delta,S,R} (z_0)
\right\rangle  ~,
\label{eq:ConformalTriangle3pt}
\end{equation}
the labels $\Delta$, $S$ and $R$ corresponding to the scaling dimension, the spin and the R-charge of the operator $\mathcal{O}_{\Delta,S,R}$, respectively.
The functions $\Omega_{\Delta,S,R}$ should form a complete basis in the vector space of a non-compact representation of the superconformal group $\mathrm{OSP}(2 \vert 4)$.
Accordingly, we formally write the completeness relation of the eigenfunctions as 
\begin{equation}
\delta^{(2)}\left(\bar{\theta}_{13}\right) \delta^{(3)}\left( x_{13} \right) \cdot
\delta^{(2)}\left(\theta_{24}\right) \delta^{(3)}\left( x_{24} \right)
=
\sumint{\Delta,S} ~~~~~~
\int \dd z_0 ~
\bar{\Omega}_{\Delta,S,R} {\scriptstyle (z_1, z_2 ; z_0) } \;
\Omega_{\Delta,S,R} {\scriptstyle (z_3, z_4 ; z_0) }
\label{eq:CompletenessRelation}
\end{equation}
and we can insert the resolution of unity into supergraphs like \eqref{eq:0MagnonPertExpansion_2} and replace the graph-builders with their eigenvalues.

For the zero-magnon case at hand, we require the superconformal triangle $\Omega_{\Delta,S,R} {\scriptstyle (z_1, z_2 ; z_0) }$ to be the three-point function of $\mathcal{O}_1 = \Phi_1^\dagger$, $\mathcal{O}_2 = \Phi_3$, both with scaling dimension $\Delta_\Phi = \frac{1}{2}$, and the exchanged operator $\mathcal{O}_{\Delta,S,0}$, which is required to have zero R-charge.
For this case, the superconformal triangles were found in \cite{Chang:2021fmd}, and for the spinless case $S=0$ they read
\begin{equation}
\Omega_{\Delta,0,0} {\scriptstyle (z_1, z_2 ; z_0)}
=
\frac{C_{\Phi_1^\dagger \Phi_3 \mathcal{O}}}
{\left[x_{2\bar{1}}^2 \right]^{\Delta_\Phi - \frac{\Delta}{2}} 
\left[x_{0\bar{1}}^2 \right]^{\frac{\Delta}{2}}
\left[x_{2\bar{0}}^2 \right]^{\frac{\Delta}{2}}} ~.
\end{equation}
We can denote the eigenvalues of the zero-magnon graph-building kernels in \eqref{eq:0MagnonPertExpansion_2} as
\begin{subequations}
\begin{align}
\int 
\dd^5 \bar{z}_1
\dd^5 z_2 ~
%\dd^3x_1\, \dd^2\bar{\theta}_1\, % \dd^2\theta_1\, \theta_1^2\;
%\dd^3x_2\, \dd^2\theta_2 \, % \dd^2\bar{\theta}_2\, \bar{\theta}_2^2\;
\Omega_{\Delta,S,0} {\scriptstyle (z_1, z_2 ; z_0)}
\left[ \mathbbm{H} \circ \bar{\mathbbm{H}} \right] {\scriptstyle (z_1, z_2 ; z_3, z_4) }
&=
E_{0} (\Delta, S)^2
\cdot \Omega_{\Delta,S,0} {\scriptstyle (z_3, z_4 ; z_0)} ~, \\
\int 
\dd^5 \bar{z}_1
\dd^5 z_2 ~
\Omega_{\Delta,S,R} {\scriptstyle (z_1, z_2 ; z_0)}
\left[ \mathbbm{H} \circ \mathbbm{P} \right] {\scriptstyle (z_1, z_2 ; z_3, z_4) }
&=
E_{0} (\Delta, S)
\cdot \Omega_{\Delta,S,0} {\scriptstyle (z_3, z_4 ; z_0)} ~.
\end{align}
\label{eq:0Magnon_ev_Omega}%
\end{subequations}
We introduced the shorthand for the chiral and anti-chiral superspace integration measure $\dd^5 z = \dd^3 x\; \dd^2 \theta\, \dd^2 \bar{\theta} ~ \bar{\theta}^2$ and $\dd^5 \bar{z} = \dd^3 x\; \dd^2 \theta\, \dd^2 \bar{\theta} ~ \theta^2$, respectively.
The computation of the eigenvalue can be facilitated by considering the limit $x_0 \rightarrow \infty$ and $\theta_0, \bar{\theta}_0 =0$, which can be inverted by superconformal transformations.
In the spinless case we study the operator $\mathcal{O}_{\Delta,0,0} = \mathrm{tr} \left[ \Phi_1 \Phi_3^\dagger \right]$ with scaling dimension $\Delta = \Delta_0 + \gamma = 1 + \gamma$ and the asymptotic expression for the corresponding eigenfunction gives
\begin{equation}
\Omega_{\Delta,0,0} (z_1, z_2 ; z_0)
\stackrel{\substack{x_0 \rightarrow \infty \\ \theta_0, \bar{\theta}_0 = 0 }}{\sim}
\Psi_{\frac{1}{2}-\frac{\Delta}{2}} (z_1 , z_2)
:=
\adjincludegraphics[valign=c,scale=1]{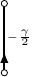} ~.
\label{eq:0Magnon_Omega_scaling}
\end{equation}
The calculation of the eigenvalue $E_{0} (\Delta) = E_{0} (\Delta, 0)$ on $\Psi_{\frac{1}{2}-\frac{\Delta}{2}} (z_1 , z_2)$ of $\mathbbm{H} \circ \bar{\mathbbm{H}}$ and $\mathbbm{H} \circ \mathbbm{P}$ is performed in appendix \ref{appsub:ZeroMagnonOperatorDiag} by the chain relations that are established in section \ref{subsec:SuperChainRelations}.
We find 
\begin{equation}
E_{0} (\Delta)
=
\frac{4 \pi ^4}{(\Delta - 1 ) \Delta } ~.
\label{eq:0Magnon_ev_spinless}
\end{equation}

Now we can return to the four-point function \eqref{eq:0MagnonPertExpansion_2} and insert unity in the form of the completeness relation \eqref{eq:CompletenessRelation}.
Then the action of $\mathbbm{H} \circ \bar{\mathbbm{H}}$ and $\mathbbm{H} \circ \mathbbm{P}$ on the eigenfunctions will give the eigenvalues by \eqref{eq:0Magnon_ev_Omega} and we can sum \eqref{eq:0MagnonPertExpansion_2} in a geometric series.
Hence, we obtain the representation 
\begin{equation}
\begin{split}
&\left\langle 
\mathrm{tr}
\left[
\Phi_1 (z_1)
\Phi_3^\dagger (z_2)
\right]
\mathrm{tr}
\left[
\Phi_1^\dagger (z_3)
\Phi_3 (z_4)
\right]
\right\rangle \\
&=
\adjincludegraphics[valign=c,scale=1]{pictures/2ptfctn/zeromagnon/diags/4ptfunction_deco0_prime.pdf} 
\circ
\left[
\sumint{\Delta,S} ~~~~~~~~
\frac{1}{1 - \xi^2 E_{0} (\Delta, S)}
\int \dd z_0 ~
\bar{\Omega}_{\Delta,S,0} {\scriptstyle (z_1', z_2' ; z_0) }\;
\Omega_{\Delta,S,R} {\scriptstyle (z_3, z_4 ; z_0) }
\right]
 ~.
\end{split}
\label{eq:0MagnonPertExpansion_3}
\end{equation}
In this form, we can observe that poles arise whenever the denominator vanishes.
This signals the exchange of the operator $\mathcal{O}_{\Delta,S,0}$.

Rather than the full four-point function, we are interested in the anomalous dimension of the exchanged operators, which amounts to solving the equation
\begin{equation}
1 
=
\xi^2 \cdot
E_{0} (\Delta, S)
\label{eq:0Magnon_polecondition}
\end{equation}
for the scaling dimension $\Delta$. 
The physical poles of $E_{0} (\Delta, S)$ in $\Delta$ signal the classical scaling dimension of an exchanged operator because the divergence can counterbalance the weak-coupling limit $\xi \rightarrow 0$ for \eqref{eq:0Magnon_polecondition} to hold.
In our case \eqref{eq:0Magnon_ev_spinless}, we find two poles at $\Delta_0 = 0$ and $\Delta_0 = 1$ and inverting \eqref{eq:0Magnon_polecondition} can be done analytically.
Finally, we obtain the exact scaling dimension of the operator $\mathrm{tr} \left[ \Phi_1 \Phi_3^\dagger \right]$ to the solution corresponding to $\Delta_0 = 1$, which is
\begin{equation}
\Delta
=
1 + \gamma
=
1 + \frac{1}{2} \left( -1 + \sqrt{ 1 + 16 \pi ^4 \xi^2}\right) ~.
\label{eq:0Magnon_result_scalingDim}
\end{equation}
Therefore, we can confirm the result for the dispersion relation of the length-2 operator, which was obtained by Bethe ansatz in \cite{Caetano:2016ydc} and by re-summation of ladder diagrams in \cite{Bak:2009tq} for the undeformed ABJM theory.

\subsection{Two-Magnon case}
\label{subsec:TwoMagnonCase}
After the successful diagonalization of the graph-builders appearing in the zero-magnon case in section \ref{subsec:ZeroMagnonCase}, we consider the supergraph expansion of another correlation function, which is
\begin{equation}
\begin{split}
& \left\langle 
\mathrm{tr}
\left[
\Phi_2 (z_1)
\Phi_1 (z_1)
\Phi_2 (z_2)
\Phi_1 (z_2)
\right]
\mathrm{tr}
\left[
\Phi_1^\dagger (z_3)
\Phi_2^\dagger (z_3)
\Phi_1^\dagger (z_4)
\Phi_2^\dagger (z_4)
\right]
\right\rangle \\
= &
\adjincludegraphics[valign=c,scale=1]{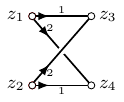}
+
\xi^4
\adjincludegraphics[valign=c,scale=1]{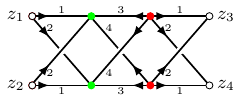}
+
\xi^8
\adjincludegraphics[valign=c,scale=1]{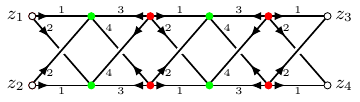} \\
& + \cdots + (z_3 \leftrightarrow z_4) ~.
\end{split}
\label{eq:2MagnonPertExpansion}
\end{equation}
We call the study of this correlation function the two-magnon case because there are two strands of propagators composed by alternating flavor-two and flavor-four superpropagators.
They wind around the cylindrical supergraphs, but unlike in the zero-magnon case \eqref{eq:0MagnonPertExpansion}, they are not closed, but rather start and end in the external operators of the correlation function.
At this point, one might wonder if there is a one-magnon case, much like in the bi-scalar fishnet theory \cite{Gromov:2018hut}.
However, the alternating chirality of the vertices in the horizontal direction rules out the corresponding supergraphs.
Furthermore, we could have considered the correlation function with flavors $(\mathrm{1} , \mathrm{2}) \leftrightarrow (\mathrm{3} , \mathrm{4})$ exchanged, which would give an equivalent derivation of the scaling dimensions.

In \eqref{eq:2MagnonPertExpansion}, we can identify the graph-builders
\begin{equation}
\begin{aligned}[c]
\mathbbm{H}
=
\adjincludegraphics[valign=c,scale=1]{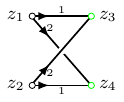} ~,
\end{aligned}
\qquad\qquad
\begin{aligned}[c]
\bar{\mathbbm{H}}
=
\adjincludegraphics[valign=c,scale=1]{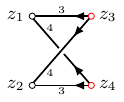} ~,
\end{aligned}
\begin{aligned}[c]
\qquad \qquad
\mathbbm{P}
=
\adjincludegraphics[valign=c,scale=1]{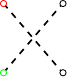} ~,
\end{aligned}
\label{eq:2MagnonGraphBuilder}
\end{equation}
and rewrite the correlation function \eqref{eq:2MagnonPertExpansion} as
\begin{equation}
 \left\langle 
\mathrm{tr}
\left[
\Phi_2 (z_1)
\Phi_1 (z_1)
\Phi_2 (z_2)
\Phi_1 (z_2)
\right]
\mathrm{tr}
\left[
\Phi_1^\dagger (z_3)
\Phi_2^\dagger (z_3)
\Phi_1^\dagger (z_4)
\Phi_2^\dagger (z_4)
\right]
\right\rangle
=
\frac{\mathbbm{H}
\circ
(1 + \mathbbm{P})}{1 - \xi^4 \mathbbm{H} \circ \bar{\mathbbm{H}} } ~.
\label{eq:2MagnonPertExpansion_2}
\end{equation}
Unfortunately, we are unaware of the precise form of the conformal triangle \eqref{eq:ConformalTriangle3pt} for the operators $\mathcal{O}_1 = \mathcal{O}_2 = \Phi_4 \Phi_3$.
However, since both operators are charged under the R-symmetry \eqref{eq:U1_intakt_transformation} with charge $R_{1} = R_{2} = \frac{1}{2} + \frac{1}{2} = 1$, the superconformal triangle corresponds to an operator $\mathcal{O}_{\Delta , S , R}$ with non-trivial R-charge $R = -2$.
We want to employ the same trick as in in the zero-magnon case \eqref{eq:0Magnon_Omega_scaling} and amputate the superspace point $z_0$ where the operator $\mathcal{O}_{\Delta , S , R}$ sits.
Therefore, we consider again the limit $x_0 \rightarrow \infty$ and set the fermionic coordinates to zero, $\theta_0 , \bar{\theta}_0 = 0$.
The two-point function that we should obtain will carry R-charge $-2$, hence, regarding \eqref{eq:FeynmanRules_AuxTwoPointFctns}, we assume the eigenfunction to be
\begin{equation}
\Omega_{\Delta,0,-2} (z_1, z_2 ; z_0)
\stackrel{\substack{ x_0 \rightarrow \infty \\ \theta_0, \bar{\theta}_0 = 0 }}{\sim}
\Psi_{\frac{1}{2}-\frac{\Delta}{2}} (z_1 , z_2)
:=
\frac{\theta_{12}^2}{\left[x_{12}^2 \right]^{\frac{1}{2}-\frac{\Delta}{2}}} 
= ~
\adjincludegraphics[valign=c,scale=1]{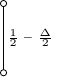} 
\label{eq:2Magnon_Omega_scaling}
\end{equation}
for the spinless case $S=0$.
To find the exponent of $\Psi_{u}$, we had to use the more general relation for its scaling dimension $\Delta_{\Psi_{u}} = 1 - 2u = \Delta - \Delta_1 - \Delta_2 + R_1 + R_2$ also considering the R-charges of the operators in $\Omega_{\Delta,0,R} (z_1, z_2 ; z_0)$.

\begin{figure}[h]
\includegraphics[scale=0.7]{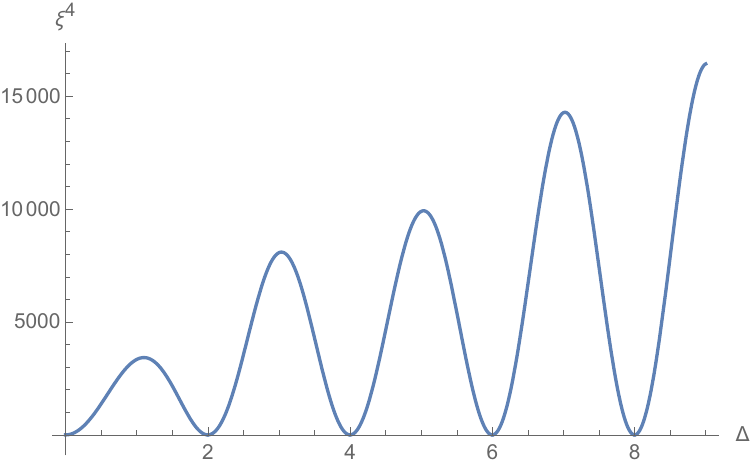} 
\centering
\caption{The plot shows the function $\xi^4 = E_2 (\Delta)^{-2}$ for $\Delta \in [0,9]$. We observe that the zeros $\Delta_0$ occur at $0$, $2$, $4$, $6$, $8$, ... . Unfortunately, the approximations in \eqref{eq:DeltaApproximatinos} are of too low order to be accurate in the shown range of the coupling. Each operator's scaling dimension will collide with the neighboring operators at a local maximum for increasing coupling. The critical value for the coupling increases with the classical dimension of the operators.}
\label{fig:Plot_xiofDelta}
\end{figure}

Next, the diagonalization of the graph-builder $\mathbbm{H}$ on the eigenfunction \eqref{eq:2Magnon_Omega_scaling} can be carried out by doing the superspace integration explicitly, as presented in appendix \ref{eq:2MagnonOperator}. 
The calculation presented there works analogously for the conjugated graph-builder $\bar{\mathbbm{H}}$ with the same eigenvalue, $\Psi_{u}^\dagger \circ \bar{\mathbbm{H}} = \mathrm{E}_2 (u) \cdot \Psi_u$.
Hence, we find the eigenvalue equation
\begin{equation}
\Psi_u \circ 
\left(\mathbbm{H} \circ \bar{\mathbbm{H}}\right) 
= 
\mathrm{E}_2 (u)^2 \cdot \Psi_u
\end{equation}
where the eigenvalue upon action on the eigenfunction is $E_2 (\Delta) = \mathrm{E}_2 (\tfrac{1}{2}-\tfrac{\Delta}{2})$ and from \eqref{eq:2Magnon_Ifactor_intrepr} and \eqref{eq:2Magnon_ev_mathrmE} we find
\begin{equation}
E_2 (\Delta)
=
\frac{\csc (\pi ( \frac{\Delta }{2} + 1 ))\, \Gamma ( \frac{\Delta}{2} + 1 )}
{32 \sqrt{\pi }\, \Gamma ( \frac{\Delta}{2} + \frac{3}{2} )}
-
\frac{\, _3F_2\left(1 , 1 , \frac{\Delta}{2} + \frac{3}{2} ; \frac{\Delta}{2} + 2 , \frac{\Delta}{2} + \frac{5}{2} ; 1 \right)}
{16 \pi ^2 \left( \Delta + 3 \right) \left(\frac{\Delta }{2}+1\right)} ~.
\label{eq:2Magnon_ev_3F2}
\end{equation}
We observe that the action of the graph-builder $\mathbbm{P}$ is trivial, since $\Psi_u$ is symmetric under exchange of the external points, $\Psi_u \circ \mathbbm{P} = \Psi_u$.
We find that the condition of \eqref{eq:2MagnonPertExpansion_2} for having a pole relates the coupling with the square of $E_2 (\Delta)$,
\begin{equation}
1 
=
\xi^4 
E_2 (\Delta)^2 ~.
\end{equation}
In the weak-coupling limit, the function $E_2 (\Delta)^2$ has to have poles at the values of classical scaling dimensions $\Delta_0$ of the exchanged operators.
We find them at the even, positive integers, including zero, $\Delta_0 \in 2\mathbbm{N}_0$, see figure \ref{fig:Plot_xiofDelta}.
Contrary to the zero-magnon case, we cannot find the inverse function of $E_2 (\Delta)^{-2}$ to access the anomalous dimensions of the exchanged operators analytically.
However, we can expand the function $E_2 (\Delta)^{-2}$ around the zeros $\Delta_0$ and invert the series perturbatively, where we find two solutions for each branch, related by $\xi^2 \rightarrow - \xi^2$.
The inversion can be done to arbitrary order in the coupling.
For the lowest three scaling dimensions, up to eight loops and after rescaling $\xi \rightarrow \pi \cdot \xi$ we find
\begin{subequations}
\begin{align}
\begin{split}
\Delta^{(2)}
~=~ &
2 
\pm \frac{\xi^2}{12} 
- \frac{\xi^4}{576} 
\pm \frac{18 \pi ^2-97}{248832} \xi^6 \\
& +\frac{ 3803 - 2268 \zeta_3 + 18 \pi ^2 (\log (4096)-19)}{35831808} \xi^8
+\mathcal{O}(\xi^{10}) ~,
\end{split} \\
\begin{split}
\Delta^{(4)}
~=~  &
4
\pm\frac{4 \xi^2}{15}
-\frac{\xi^4}{7200}
\pm \frac{28800 \pi ^2-191191}{777600000} \xi^6 \\
& \frac{ 191678057 - 145152000 \zeta_3 + 28800 \pi ^2 (480 \log (2)-421)}{5598720000000} \xi^8 
+\mathcal{O}(\xi^{10}) ~,
\end{split} \\
\begin{split}
\Delta^{(6)}
~=~ &
6
\pm \frac{2 \xi^2}{35}
-\frac{4 \xi^4}{2940}
\pm \frac{ 352800 \pi ^2 - 2244421}{15126300000} \xi^6 \\
& +\frac{2972114029 - 2074464000 \zeta_3 + 117600 \pi ^2 (1680 \log (2)-1801)}{148237740000000} \xi^8
+\mathcal{O}(\xi^{10}) .
\end{split}
\end{align}
\label{eq:DeltaApproximatinos}%
\end{subequations}
The scaling dimension $\Delta^{(2)}$ corresponds to the exchanged operator $\mathrm{tr}\left[ \Phi_3^\dagger \Phi_4^\dagger \Phi_3^\dagger \Phi_4^\dagger \right]$, which matches with the classical values of the charges $\Delta^{(2)} = 2$ and $R = -2$.
The other solutions with higher classical scaling dimensions corresponds to insertions of bosonic space derivatives, like e.\,g.\ $\square \; \mathrm{tr}\left[ \Phi_3^\dagger \Phi_4^\dagger \Phi_3^\dagger \Phi_4^\dagger \right]$.

\section{Conclusions and Outlook}
We summarize our results.
After performing a $\beta$-deformation on part of the R-symmetry of ABJM theory and taking a double-scaling limit of the 't Hooft coupling and the deformation parameter, we obtained the three-dimensional $\mathcal{N} = 2$ superfishnet theory \eqref{eq:Action_superfishnet_Z_W}, or equivalently \eqref{eq:Action_superfishnet_Phi}.
Additionally, we proposed a non-local deformation thereof, presented in \eqref{eq:Action_genSuperfishnet_Phi}.
Despite containing many interaction vertices in the spacetime action, the superspace Feynman graphs have a very regular fishnet-like appearance, which motivates us to study the theory in a supersymmetry covariant way.
For this sake, in section \ref{sec:SuperfishnetTheory} we established the super Feynman rules and beneficial auxiliary relations, the super chain relations and super x-unity, suited for evaluating the supergraphs.
These relations can be understood as the superspace analogs of relations that manifest the integrability in bi-scalar fishnet theory.
Therefore, we continued with applications of our superspace formalism, inspired by calculations in the bi-scalar fishnet theory: 
We performed the ``zeroth-order integrability check'' of Zamolodchikov \cite{Zamolodchikov:1980mb}, and we obtained the critical coupling of the superfishnet theory, as well as the generalized superfishnet theory by the method of inversion relations.
Moreover, by studying two types of four-point correlation functions, we could show that they are constructed by consecutive action of superspace graph-builders.
In the appendix \ref{subsec:DiagonalizationOfTheSuperGraphBuildingOperators}, we show how to diagonalize them on superspace three-point functions, the superconformal triangles.
The graph-builder's eigenvalues give us access to the all-loop scaling dimensions of length-two and -four operators, with the results \eqref{eq:0Magnon_result_scalingDim} and \eqref{eq:DeltaApproximatinos}.
Furthermore, the superspace diagonalization technique to find all-loop anomalous dimensions was also applied to the four-dimensional $\mathcal{N} = 1$ super brick wall theory in appendix \ref{subsubsec:ZeroMagnonSBW}, where we find the known results for the supersymmetric dynamical fishnet theory.

Our work suggests many directions for further research.
First, the elephant in the room is the explicit construction of a super star-triangle relations.
That would be a manifest proof of integrability on the supergraph level, since it would allow the construction of an R-matrix similar to the bosonic spacetime case \cite{Chicherin:2012yn,Gromov:2017cja}.
In the one-dimensional supersymmetric case of the superalgebra superalgebra $\mathfrak{sl}(2 \vert 1)$ this was achieved in \cite{Derkachov:2001sx,Derkachov:2005hx,Belitsky:2006cp}.
Secondly, one might extend our analysis of the zero-magnon case to include spin.
Therefore, the chain relations have to be generalized to superpropagators with spin, which works analogously as in the bosonic case \cite{Derkachov:2021rrf}.
This would allow us to calculate the eigenvalues with spin, since the spinning conformal triangles are known due to \cite{Chang:2021fmd}.
Then, one can access the OPE coefficients and the full superconformal four-point function.
Furthermore, the explicit construction of the R-charged superconformal triangles would allow this analysis to be extended to the two-magnon case.
Finally, one can extend the superspace analysis of the superfishnet theory to the many investigations of the bi-scalar fishnet theory and its relatives, presumably requiring the super star-triangles relation, or one might study the spectral problem perturbatively \cite{Ipsen:2018fmu,Ahn:2020zly,Ahn:2021emp,Ahn:2022snr}.

Further possible investigations are the determination of super Basso-Dixon diagrams \cite{Basso:2017jwq,Derkachov:2020zvv,Derkachov:2021rrf,Basso:2021omx,Loebbert:2024fsj}, the study of correlation functions of operators with higher length by diagonalization of mirror super graph-builders and TBA \cite{Basso:2018agi,Basso:2019xay}, the construction of the holographic dual of the superfishnet theory in analogy with \cite{Gromov:2019aku,Gromov:2019bsj,Gromov:2019jfh,Gromov:2021ahm,iakhibbaev2023generalising} and the one of the Yangian of the superconformal algebra and, finally, the repetition of the Yangian bootstrap program \cite{Chicherin:2017frs,Corcoran:2021gda} for supergraphs as well as their related supergeometries \cite{Duhr:2022pch,Duhr:2023eld,Duhr:2024hjf}.

\paragraph{Acknowledgements}
The author is grateful for the support, collaboration and many discussions with Matthias Staudacher.
Furthermore, the author would like to thank Changrim Ahn for collaboration and discussing related subjects.
The author is thankful to have benefited from discussions with Changrim Ahn, Gwena\"{e}l Ferrando, Nikolay Gromov,  Vladimir Kazakov, Enrico Olivucci, Giulia Peveri, Lorenzo Di Pietro and Matthias Staudacher.
Special thanks to Matthias Staudacher and Changrim Ahn for very careful readings of the manuskript.
This work is funded by the Deutsche Forschungsgemeinschaft (DFG, German Research Foundation) - Projektnummer 417533893/GRK2575 “Rethinking Quantum Field Theory”.

%%%%%%%%%%%%%%%%%%%%%%%%%%%%%%%%%%%%%%%%%%%%%%%%%%%%%%%%%%%%%%%%%%%%%%%%%%%%%%%%%%%%%%%%%%%%%%%%%%%%%%%%%%%%%%%%%%%%%%%%%%%%%%%%%%%%%%%%%%%%%
\appendix

\section{Notations}
\label{sec:Notations}

\subsection{Spinor algebra}
We follow the notation of \cite{Benna:2008zy} for the spinor conventions.
The antisymmetric $\varepsilon_{\alpha \beta}$ has the components $\varepsilon^{12} = - \varepsilon_{12} = 1$, such that $\varepsilon^{\alpha \beta} \varepsilon_{\beta \gamma} = \delta^\alpha_\gamma$.
The three-dimensional gamma matrices are $\gamma^{\mu\; \beta}_{~\alpha} = \left( \I \sigma^2 , \sigma^1 , \sigma^3  \right)$ and they satisfy the Clifford algebra $\gamma^{\mu\; \beta}_{~\alpha} \gamma^{\nu\; \gamma}_{~\beta} = g^{\mu\nu} \delta^\gamma_\alpha + \varepsilon^{\mu\nu\rho} \gamma^{~~ \gamma}_{\rho\,\alpha}$ with the spacetime metric $g^{\mu\nu} = \mathrm{diag}(-1, 1, 1)$.
Lowering an index gives $\gamma_{\alpha\beta}^\mu = \varepsilon_{\beta \delta} \gamma^{\mu\; \delta}_{~\alpha} = \left( - \mathbbm{1} , - \sigma^3 , \sigma^1 \right)$, which is symmetric in the spinor indices $\gamma_{\alpha\beta}^\mu = \gamma_{\beta\alpha}^\mu$.
Furthermore, the trace $\gamma^{\mu\; \alpha}_{~\alpha} = 0$ vanishes.

Spinor indices of anti-commuting fermions and Gra\ss mann numbers are raised and lowered with the antisymmetric epsilon tensor according to 
\begin{equation}
\begin{aligned}[c]
\psi^\alpha = \varepsilon^{\alpha \beta} \psi_{\beta} ~,\\
\psi_\alpha = \varepsilon_{\alpha \beta} \psi^{\beta} ~,
\end{aligned}
\qquad\qquad
\begin{aligned}[c]
\bar{\psi}^{\alpha} = \varepsilon^{\alpha \beta} \bar{\psi}_{\beta} ~,\\
\bar{\psi}_{\alpha} = \varepsilon_{\alpha \beta} \bar{\psi}^{\beta} ~,
\end{aligned}
\end{equation}
and the spinor bilinears are 
\begin{equation}
\begin{aligned}[c]
\psi \chi & = \psi^\alpha \chi_\alpha = - \psi_\alpha \chi^\alpha ~,\\
\bar{\psi} \bar{\chi} & = \bar{\psi}^{\alpha} \bar{\chi}_{\alpha} = - \bar{\psi}_{\alpha} \bar{\chi}^{\alpha} ~,
\end{aligned}
\qquad\qquad
\begin{aligned}[c]
\psi \gamma^\mu \bar{\chi} = \psi^\alpha \gamma^\mu_{\alpha \beta} \bar{\chi}^{\beta} ~.
\end{aligned}
\end{equation}
The squares of spinorial Gra\ss mann numbers, for example, are denoted by $\theta^2 := \theta^\alpha \theta_\alpha$ and $\bar{\theta}^2 := \bar{\theta}^\alpha \bar{\theta}_\alpha$ and we have the frequently used relations
\begin{equation}
\begin{aligned}[c]
\theta^\alpha \theta^\beta &= - \frac{1}{2} \varepsilon^{\alpha \beta} \theta^2 ,\\
\bar{\theta}^{\alpha} \bar{\theta}^{\beta} &= - \frac{1}{2} \varepsilon^{\alpha \beta} \bar{\theta}^2 .
\end{aligned}
\qquad\qquad
\begin{aligned}[c]
\theta \gamma^\mu \bar{\theta}\; \theta \gamma^\nu \bar{\theta} = \frac{1}{2} \theta^2 \bar{\theta}^2 g^{\mu\nu} ,
\end{aligned}
\label{eq:ThetaSquare_ThetaCube}
\end{equation}
Other relations, like the more general Fierz identities, can be found in the appendix of \cite{Benna:2008zy}.

\subsection{Berezin integral}
\label{sec:BerezinIntegral}
The quadratic component in the Gra\ss mann spinors is picked out by Berezin integration over the fermionic part of superspace, which also implies
\begin{equation}
\int \dd^2 \theta \; \theta^2  =  1
~~ \mathrm{and} ~~
\int \dd^2 \bar{\theta} \; \bar{\theta}^2  =  1~ .
\label{eq:BerezinIntegral}
\end{equation}
Since terms cubic in the Gra\ss mann spinors vanish due to the anti-commutativity, squares of the Gra\ss mann spinors act as delta distributions on fermionic superspace.
We denote them by $\delta^{\left(2\right)} \left( \theta \right) = \theta^2$ and $\delta^{\left(2\right)} \left( \bar{\theta} \right) = \bar{\theta}^2$.
Regarding the mass dimension, Gra\ss mann spinors have half the dimension assigned to bosonic coordinates, i.\,e.\ $\left[ \theta^\alpha \right] = \left[ \bar{\theta}^{\dot{\alpha}} \right] = \frac{\left[ x^\mu \right]}{2} = -\frac{1}{2}$.
By \eqref{eq:BerezinIntegral}, the measure of the Berezin integral has the inverse mass dimensions of a Gra\ss mann square, namely $\left[ \dd^2 \theta \right] = \left[ \dd^2 \bar{\theta} \right] = 1$.

According to \cite{Benna:2008zy}, the covariant super derivatives and supersymmetry generators read
\begin{subequations}
\begin{align}
D_\alpha 
=
\partial_\alpha + \I\, \gamma^\mu_{\alpha \beta} \bar{\theta}^{\beta} \partial_\mu ~,
\hspace{1cm}
\bar{D}_{\dot{\alpha}} 
=
- \bar{\partial}_{\alpha} - \I\, \theta^\beta \gamma^\mu_{\beta \alpha} \partial_\mu  ~, \label{eq:SuperCovariantDerivatives}%
\\
Q_\alpha 
=
\partial_\alpha - \I\, \gamma^\mu_{\alpha \beta} \bar{\theta}^{\beta} \partial_\mu ~,
\hspace{1cm}
\bar{Q}_{\alpha} 
=
- \bar{\partial}_{\alpha} + \I\, \theta^\beta \gamma^\mu_{\beta \alpha} \partial_\mu  ~,
\end{align}
\end{subequations}
with $\partial_\alpha = \frac{\partial}{\partial \theta^\alpha}$, $\bar{\partial}_{\dot{\alpha}} = \frac{\partial}{\partial \bar{\theta}^{\dot{\alpha}}}$ and $\partial_\mu = \frac{\partial}{\partial x^\mu}$.
Integration and derivation for Gra\ss mann numbers are equivalent, concretely
\begin{subequations}
\begin{align}
\int \dd^2 \theta \; f(\theta)  = \left[ - \frac{1}{4} D^2 f(\theta) \right]_{\theta = 0} ~, &
\hspace{1cm}
\int \dd^2 \bar{\theta} \; f(\bar{\theta})  = \left[ - \frac{1}{4} \bar{D}^2 f(\bar{\theta}) \right]_{\bar{\theta} = 0}  ~,\\
\int \dd^2 \theta\, \dd^2 \bar{\theta} \; f(\theta,\bar{\theta}) 
& = \left[ \frac{1}{16} D^2 \bar{D}^2 f(\theta,\bar{\theta}) \right]_{\theta, \bar{\theta} = 0 } ~.
\end{align}
\end{subequations}

\section{Details of the super-integral calculations}

\subsection{An auxiliary relation for the construction of super x-unity}
\label{subsec:auxRelForXUnity}
We present the derivation of \eqref{eq:3ptFctnOnePointIntegrated} by direct calculation, using the Feynman rules of \ref{sec:SuperfishnetTheory}.
The diagram reads
\begin{equation}
\I \int \dd^3x_0\; \dd^2\theta_0\,\dd^2\bar{\theta}_0 \;\delta^{(2)}(\theta_0)
\frac{1}{\left[ x_{2\bar{0}} \right]^{u_2}}
\frac{1}{\left[ x_{3\bar{0}} \right]^{1-u_1-u_2}} ~
\I \int \dd^3x_1\; \dd^2\theta_1\,\dd^2\bar{\theta}_1 \;\delta^{(2)}(\bar{\theta}_1)
\frac{1}{\left[ x_{1\bar{0}} \right]^{u_1}} \vert_{\bar{\theta}_{2,3} = 0} ~.
\label{eq:AuxRelDiagram}
\end{equation}
The fermionic part of the superspace integral over $z_1$ can be performed, and we find
\begin{equation}
\I \int \dd^3x_1\; \dd^2\theta_1\,\dd^2\bar{\theta}_1 \;\delta^{(2)}(\bar{\theta}_1)
\frac{1}{\left[ x_{1\bar{0}} \right]^{u_1}}
=
- 4 u_1 (\tfrac{1}{2} - u_1)\;
\bar{\theta}_0\;
\I \int \dd^3x_1\;
\frac{1}{\left[ x_{10} \right]^{u_1 + 1}} ~.
\label{eq:AuxRel_1}
\end{equation}
Here we used the relation $\square_{1} \frac{1}{\left[ x_{12}^2 \right]^u} = - 4u (\frac{3}{2} - u -1) \frac{1}{\left[ x_{12}^2 \right]^{u+1}}$.
Plugging \eqref{eq:AuxRel_1} into \eqref{eq:AuxRelDiagram}, we can perform the $z_0$ superspace integral.
The fermionic integrations evaluate the fermionic delta functions $\theta_0^2$ and $\bar{\theta}_0^2$, and the bosonic integral is performed by the bosonic star-triangle relation in three dimensions, see (3.3) in \cite{Kade:2023xet}.
We obtain for \eqref{eq:AuxRelDiagram}
\begin{equation}
- 4 u_1 (\tfrac{1}{2} - u_1) r(u_2, 2-u_1-u_2, u_1 +1)
\int \dd^3x_1\; 
\frac{1}{\left[ x_{13} \right]^{\frac{3}{2} - u_2}}
\frac{1}{\left[ x_{23} \right]^{\frac{1}{2} - u_1}}
\frac{1}{\left[ x_{12} \right]^{u_1 + u_2 - \frac{1}{2}}}
\end{equation}
Next, we use the bosonic chain relation to evaluate the integration over $x_1$, see (3.7) in \cite{Kade:2023xet}, to find the dependence on the external points vanishing and \eqref{eq:AuxRelDiagram} turning into a factor
\begin{equation}
- 4 u_1 (\tfrac{1}{2} - u_1)\;
r(u_2, 2-u_1-u_2, u_1 +1)\;
r(2-u_1, \tfrac{3}{2}-u_2, u_1 + u_2 - \tfrac{1}{2}) ~.
\end{equation}
Finally, the result of \eqref{eq:3ptFctnOnePointIntegrated} can be obtained by using the explicit form of the $r$-factor \eqref{eq:RfactorAfactor} and the functional relation of the $\Gamma$-function.

\subsection{Diagonalization of the supergraph-building operators}
\label{subsec:DiagonalizationOfTheSuperGraphBuildingOperators}

\subsubsection{Zero-Magnon operator}
\label{appsub:ZeroMagnonOperatorDiag}
We show that the operators $\mathbbm{H} \circ \bar{\mathbbm{H}}$ and $\mathbbm{H} \circ \mathbbm{P}$ have the eigenfunctions $\Psi_u$.
The generalized 0-magnon graph building operator $\mathbbm{H}_{0, \boldsymbol{\delta}}$ has a valuable action on the zero-R-charge eigenfunction $\Psi_u$ that has the form of a generalized superpropagator \eqref{eq:FeynmanRules_superWeight_graphical}.
It is not an eigenfunction, however, we can use the chain rules of section \ref{subsec:SuperChainRelations} to obtain 
\begin{equation}
\begin{split}
&
\Psi_u \circ \mathbbm{H}_{0, \boldsymbol{\delta}}
=
\adjincludegraphics[valign=c,scale=1.2]{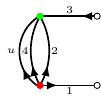} 
=
\left[
\prod_{i = 1}^4
c_{\delta_i}
\right] \cdot
\adjincludegraphics[valign=c,scale=1.2]{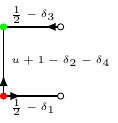}\\
=&
\left[
\prod_{i = 1}^4
c_{\delta_i}
\right]
4\I\, r (\tfrac{1}{2} - u - \delta_3 , u + 1 - \delta_2 - \delta_4, \tfrac{1}{2} - \delta_1) \cdot
\adjincludegraphics[valign=c,scale=1.2]{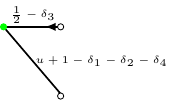} \\
=&
-
\left[
\prod_{i = 1}^4
a (1 + \delta_i)
\right]
r (\tfrac{1}{2} - u - \delta_3 , u + 1 - \delta_2 - \delta_4, \tfrac{1}{2} - \delta_1)\,
r (\tfrac{3}{2} - u, u + 1 + \delta_3, \tfrac{1}{2} - \delta_3) \cdot
\adjincludegraphics[valign=c,scale=1]{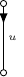}\\
=& ~\mathrm{E}_{0,\boldsymbol{\delta}} (u) \cdot \Psi_u^\dagger ~.
\end{split}
\label{eq:0MagnonPreDiag}
\end{equation}
We recall the restriction on the deformation parameters $\delta_1 + \delta_2 + \delta_3 + \delta_4 = 0$.
The factor $\mathrm{E}_{0,\boldsymbol{\delta}} (u)$ is a combination of $\Gamma$-functions involving the deformation parameters $\delta_i$ and the label of the eigenfunction $u$.
After using properties of the $\Gamma$-function, we find
\begin{equation}
\mathrm{E}_{0,\boldsymbol{\delta}} (u)
=
-\pi ^3  
\frac{\Gamma \left(\frac{1}{2}-\delta_2\right) \Gamma \left(\frac{1}{2}-\delta_4\right) \Gamma (u) \Gamma \left(-u+\delta_2+\delta_4+\frac{1}{2}\right)}
{\Gamma (\delta_2+1) \Gamma (\delta_4+1) \Gamma \left(\frac{3}{2}-u\right) \Gamma (u-\delta_2-\delta_4+1)} ~.
\end{equation}
Note that $\mathrm{E}_{0,\boldsymbol{\delta}} (u)$ only depends on $\delta_2$ and $\delta_4$.
From \eqref{eq:0MagnonPreDiag}, and its hermitian conjugate $\Psi^\dagger_u \circ \mathbbm{H}_{0, \boldsymbol{\delta}}^\dagger  = \mathrm{E}_{0,\boldsymbol{\delta}} (u) \cdot \Psi_u$, we find that $\mathbbm{H} \circ \bar{\mathbbm{H}}$ has the eigenfunction $\Psi_u$.
Furthermore, we observe that the permutation operator $\mathbbm{P}$ that permutes the start- and end-point maps $\Psi_u$ to its hermitian conjugate $\Psi_u^\dagger$.
We can combine it with and find the eigenvalue equations
\begin{subequations}
\begin{align}
\Psi_u 
\circ 
\left( \mathbbm{H} \circ \bar{\mathbbm{H}} \right)
&=
\mathrm{E}_{0,\boldsymbol{\delta}} (u)^2
\cdot \Psi_u ~, \\
\Psi_u 
\circ 
\left( \mathbbm{H} \circ \mathbbm{P} \right)
&=
\mathrm{E}_{0,\boldsymbol{\delta}} (u)
\cdot \Psi_u ~.
\end{align}
\end{subequations}
Setting the deformation parameters to zero, $\boldsymbol{\delta} = 0$, and substituting $u = \frac{1}{2} - \frac{\Delta}{2}$ gives the zero-magnon eigenvalue of the superfishnet theory
\begin{equation}
E_{0} (\Delta)
=
\mathrm{E}_{0,\boldsymbol{\delta}=0} (\tfrac{1}{2} - \tfrac{\Delta}{2})
=
-\frac{4 \pi ^4}{(1-\Delta ) \Delta } ~.
\end{equation}

\subsubsection{Two-Magnon operator}
\label{eq:2MagnonOperator}
The eigenvalue of the two-magnon graph-builder can be determined by reducing the superspace integral to the three-dimensional, bosonic kite integral \cite{Kotikov:2024yok}.
Since the kite integral is more complicated in the case of arbitrary propagator powers \cite{Grozin:2012xi}, we proceed with the calculation in the case of the undeformed superfishnet theory \eqref{eq:Action_superfishnet_Phi}.
Acting with the graph-builder $\mathbbm{H}$ of \eqref{eq:2MagnonGraphBuilder} on the eigenfunction, we find
\begin{equation}
\begin{split}
\Psi_u \circ \mathbbm{H}
&=
\adjincludegraphics[valign=c,scale=1]{pictures/2ptfctn/twomagnon/diagonalization_left/Builder_H_ev.pdf}
=
-\left[
\int 
\dd^5 z_0 \; \dd^5 z_{0'} ~
\frac{c_0}{\left[x_{0\bar{1}}^2 \right]^{\frac{1}{2}}}
\frac{c_0}{\left[x_{0'\bar{1}}^2 \right]^{\frac{1}{2}}} 
\frac{\theta_{00'}^2}{\left[x_{00'}^2 \right]^{u}} 
\frac{c_0}{\left[x_{0\bar{2}}^2 \right]^{\frac{1}{2}}}
\frac{c_0}{\left[x_{0'\bar{2}}^2 \right]^{\frac{1}{2}}} 
\right]_{\theta_{1,2} = 0} \\
&=
-c_0^4
\int 
\dd^2 \theta_0 ~
\e^{2\I \theta_0 \gamma^\mu \bar{\theta}_{12} \partial_{1,\mu}}
\int \dd^3 x_0 \; \dd^3 x_{0'}
\frac{1}{\left[x_{10}^2 \right]^{\frac{1}{2}}}
\frac{1}{\left[x_{10'}^2 \right]^{\frac{1}{2}}} 
\frac{1}{\left[x_{00'}^2 \right]^{u}} 
\frac{1}{\left[x_{20}^2 \right]^{\frac{1}{2}}}
\frac{1}{\left[x_{20'}^2 \right]^{\frac{1}{2}}} \\
&=
c_0^4 \cdot
\bar{\theta}_{12}^2 \square_1 \;
\mathsf{kite}^{(3)}
( x_{12}^2, u )  ~.
\end{split}
\label{eq:2Magnon_ev_1}
\end{equation}
Here, we have evaluated the Gra\ss mann delta function $\theta_{00'}^2$ by the $0'$-integration, which allows us to factor out the common $\theta_0$-dependent exponential with the differential operators.
The integral over bosonic space is the three-dimensional kite diagram, which can be determined as $\mathsf{kite}^{(3)}( x_{12}^2, u ) = \frac{I^{(3)}(u)}{\left[ x^2 \right]^{u-1}}$ \cite{Kotikov:2024yok}, with the function
\begin{equation}
I^{(3)}(u)
=
\frac{1}{\pi^2 \left( \frac{3}{2} - u \right) \left( u-1 \right) }
\frac{1}{32 \pi^2}
\int_1^\infty
\dd s ~
\frac{ s^{\frac{1}{2} - u} + s^{ - 2 + u } }{\sqrt{1 + s}}\;
\mathrm{log}
\left[
\frac{\sqrt{1 + s} + 1}{\sqrt{1 + s} - 1}
\right] ~.
\label{eq:2Magnon_Ifactor_intrepr}
\end{equation}
The integral representation can be expressed in terms of the hypergeometric $_3 F_2$-function at unit argument; the corresponding expression for the eigenvalue is presented in \eqref{eq:2Magnon_ev_3F2}.
The eigenvalue, corresponding to the eigenvalue equation $\Psi_u \circ \mathbbm{H} = \mathrm{E}_2 (u) \cdot \Psi_u^\dagger$, is obtained after performing the derivatives in \eqref{eq:2Magnon_ev_1} and reads
\begin{equation}
\mathrm{E}_2 (u)
=
-4 c_0^4 \cdot
\left(
u - 1 
\right)
\left(
\tfrac{3}{2} - u
\right)
I^{(3)} (u) ~.
\label{eq:2Magnon_ev_mathrmE}
\end{equation}

\subsubsection{Zero-Magnon correlation function in the super brick wall theory}
\label{subsubsec:ZeroMagnonSBW}
The assumption of the existence of a complete basis of superconformal eigenfunctions of the various graph-building operators in section \ref{subsec:TwoMagnonCase} admittedly seems ad-hoc.
Therefore, we show in this appendix that the conjectured completeness relation and the resulting anomalous dimensions agree with the literature \cite{Kazakov:2018gcy}.
However, we switch from the three-dimensional superconformal setup of the superfishnet theory to the four-dimensional superconformal brick wall theory \cite{Kade:2024ucz}.
This theory is the double-scaled $\beta$-deformation of $\mathcal{N} = 4$ SYM, which is the supersymmetric case of the so-called dynamical fishnet theory \cite{Kazakov:2018gcy}.

We use the conventions of \cite{Kade:2024ucz} for four-dimensional $\mathcal{N} = 1$ superspace and consider the supergraph expansion of the correlator
\begin{equation}
\begin{split}
&\left\langle 
\mathrm{tr}
\left[
\Phi_1 (z_1)
\Phi_1 (z_2)
\right]
\mathrm{tr}
\left[
\Phi_1^\dagger (z_3)
\Phi_1^\dagger (z_4)
\right]
\right\rangle \\
= &
\adjincludegraphics[valign=c,scale=1]{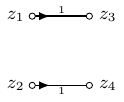}
+
\xi^4
\adjincludegraphics[valign=c,scale=1]{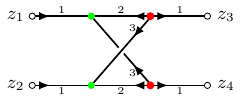}
+
\xi^8
\adjincludegraphics[valign=c,scale=1]{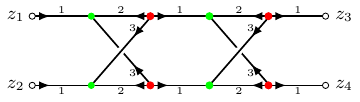} \\
&+
\xi^{12}
\adjincludegraphics[valign=c,scale=1]{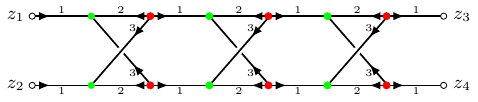} 
+
\cdots
+
\left(
z_3 
\leftrightarrow
z_4
\right) 
\end{split}
\label{eq:0MagnonPertExpansion_SBW}
\end{equation}
in the large-$\mathrm{N}$ limit.
The super brick wall theory's chiral, non-unitary, cubic vertices allow only ladder-like, cylindrical supergraphs.
Note that in \eqref{eq:0MagnonPertExpansion_SBW}, the upper-left and lower-right vertex of each rung is drawn with the wrong flavor orientation.
This is for illustrational reasons only; to be more precise, the connecting superpropagator should close after winding around the compact direction of the cylinder.

The graph-building operators in the diagrams \eqref{eq:0MagnonPertExpansion_SBW} are 
\begin{equation}
\begin{aligned}[c]
\mathbbm{H}
=
\adjincludegraphics[valign=c,scale=1]{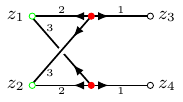} ~,
\end{aligned}
\qquad\qquad
\begin{aligned}[c]
\mathbbm{P}
=
\adjincludegraphics[valign=c,scale=1]{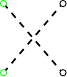} ~,
\end{aligned}
\label{eq:0MagnonGraphBuilder_SBW}
\end{equation}
and we will show that $\mathbbm{H}$ and $\mathbbm{H} \circ \mathbbm{P}$ have the same eigenvalue on the eigenfunction of our interest.
With the graph-building operators, we can formally write the correlation function \eqref{eq:0MagnonPertExpansion_SBW} as
\begin{equation}
\left\langle 
\mathrm{tr}
\left[
\Phi_1 (z_1)
\Phi_1 (z_2)
\right]
\mathrm{tr}
\left[
\Phi_1^\dagger (z_3)
\Phi_1^\dagger (z_4)
\right]
\right\rangle 
= 
\adjincludegraphics[valign=c,scale=1]{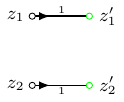}
\circ 
\left[
\frac{\left( 1 + \mathbbm{P} \right)}{1 - \xi^4 \mathbbm{H}}
\right] ~.
\label{eq:0MagnonPertExpansion_SBW_1}
\end{equation}
The eigenvalues of $\mathbbm{H}$ corresponding to the eigenfunction $\Psi_u = \tfrac{\bar{\theta}_{12}^2}{\left[ x_{12}^2 \right]^u}$ can be calculated similarly to \eqref{eq:2Magnon_ev_1}, with the difference that we are in four-dimensional $\mathcal{N} = 1$ superspace and that we have to convolute the four-dimensional kite diagram $\mathsf{kite}^{(4)} ( x^2, u ) = \frac{I^{(4)}(u)}{\left[ x^2 \right]^{u}}$ \cite{Gromov:2018hut} with the two more superpropagators.
We observe that the permutation graph-builder $\mathbbm{P}$ has a trivial action on the eigenfunction $\Psi_u$ since the eigenfunction is symmetric under the exchange of the external points.
Furthermore, the eigenfunction carries R-charge such that we expect the parameter $u$ to be related to the scaling dimension of the eigenfunction, which is as in the two-magnon case given by $\Delta_{\Psi_{u}} = \Delta - \Delta_1 - \Delta_2 + R_1 + R_2$.
Now the R-charge of $\mathcal{O}_1 = \mathcal{O}_2 = \Phi_1^\dagger$ is $R_1 = R_2 = - \frac{1}{2}$, which implies $u= 2 - \frac{\Delta}{2}$.
For the graph-builder $\mathbbm{H}$ we find the following eigenvalue equation,
\begin{equation}
\begin{split}
\Psi_u \circ \mathbbm{H} ~
&=
\adjincludegraphics[valign=c,scale=1]{pictures/2ptfctn/zeromagnon_SBW/diagonalization_left/Builder_H_SBW_ev.pdf}
=
- 4 u \left( 1 - u \right)
I^{(4)} (u) \cdot
\adjincludegraphics[valign=c,scale=1]{pictures/2ptfctn/zeromagnon_SBW/diagonalization_left/Builder_H_SBW_ev_1.pdf}\\
&=
16\, u \left( 1 - u \right)\; 
I^{(4)} (u)\;
r (2 - u , u + 1 , 1)\,
r (2 - u , u , 1) \cdot
\Psi_u ~.
\end{split}
\label{eq:0Magnon_SBW_ev_1}
\end{equation}
In the last step, we used the four-dimensional $\mathcal{N} = 1$ versions of \eqref{eq:ChainRelChiral_Aux} and \eqref{eq:ChainRelChiral}.
The eigenvalue corresponding to $\Psi_u$ is therefore 
\begin{equation}
\mathrm{E}^\mathrm{SBW}_0 (u)
=
16\, u \left( 1 - u \right) \cdot
I^{(4)} (u) \cdot
r (2 - u , u + 1 , 1)\,
r (2 - u , u , 1) ~,
\end{equation}
where the factor from the kite diagram reads \cite{Gromov:2018hut}
\begin{equation}
I^{(4)} (u) 
= 
\frac{1}{2 u-2}
\left[
\psi ^{(1)}\left(\tfrac{u-1}{2}\right)
- \psi ^{(1)}\left(\tfrac{1-u}{2}\right)
+ \psi ^{(1)}\left(\tfrac{2-u}{2}\right)
- \psi ^{(1)}\left(\tfrac{u}{2}\right)
\right] ~,
\end{equation}
with the second derivative of the logarithm of the $\Gamma$-function $\psi ^{(1)} (z) = \frac{\dd^2}{\dd z^2} \mathrm{log}\, \Gamma (z)$.
We observe again that the action of $\mathbbm{P}$ on $\Psi_u$ is trivial.
Using the functional relation of $\psi ^{(1)} (z)$ and rescaling the coupling of the super brick wall theory $\xi \rightarrow \frac{\xi}{2\pi}$, the pole condition of \eqref{eq:0MagnonPertExpansion_SBW_1}, $1 = \xi^4 \mathrm{E}^\mathrm{SBW}_0 (2 - \tfrac{\Delta}{2})$, is equivalent to the result of the dynamical fishnet theory, eq.\ (4.28) in \cite{Kazakov:2018gcy}, specified to the supersymmetric, spinless case $\omega , S = 0$ and $\Delta = 2 + 2\I \nu$ therein.

\bibliography{paper_susy_invrel}
\bibliographystyle{OurBibTeX}
\end{document}